\documentclass{jkas}

\def\farcs{\hbox{$.\mkern-4mu^{\prime\prime}$}}

\def\year{2023} 
\def\volume{56} 
\def\issue{1} 
\def\beginpage{1} 
\def\received{---} 
\def\accepted{---} 
\def\published{---} 
\date{Received \received; Accepted \accepted; Published \published}

\usepackage{ulem}

\title{%
Galaxy-Galaxy Blending in SPHEREx Survey Data
}

\author[1]{Kim Dachan}{0009-0001-3899-7198}
\author[1,$\star$]{Hyunmi Song}{0000-0002-4362-4070}
\author[2]{Yigon Kim}{0000-0001-5100-4886}
\author[2,$\star$]{Minjin Kim}{0000-0002-3560-0781}
\author[3]{Hyunjin Shim}{0000-0002-4179-2628}
\author[4]{Dohyeong Kim}{0000-0002-6925-4821}
\author[5]{Yongjung Kim}{0000-0003-1647-3286}
\author[5]{Bomee Lee}{0000-0003-1954-5046}
\author[2,6]{Jeong Hwan Lee}{0000-0003-3301-759X}
\author[5]{Woong-Seob Jeong}{0000-0002-2770-808X}
\author[5,7]{Yujin Yang}{0000-0003-3078-2763}

\affil[1]{Department of Astronomy and Space Science, Chungnam National University, 99 Daehak-ro, Yuseong-gu, Daejeon 34134, Republic of Korea}
\affil[2]{Department of Astronomy and Atmospheric Sciences, Kyungpook National University, 80 Daehak-ro, Buk-gu, Daegu 41566, Republic of Korea}
\affil[3]{Department of Earth Science Education, Kyungpook National University, 80 Daehak-ro, Buk-gu, Daegu 41566, Republic of Korea}
\affil[4]{Department of Earth Sciences, Pusan National University, Busan 46241, Republic of Korea}
\affil[5]{Korea Astronomy and Space Science Institute, Daejeon 34055, Republic of Korea}
\affil[6]{The Center for High Energy Physics, Kyungpook National University, Daegu 41566, Republic of Korea}
\affil[7]{University of Science and Technology, 217, Gajeong-ro, Yuseong-gu, Daejeon 34113, Republic of Korea}

\def\corrauthor{%
H. Song and M. Kim, \email{hmsong@cnu.ac.kr, mkim@knu.ac.kr}
}

\def\runningauthor{%
Dachan et al.
}

\def\runningtitle{%
Blending in SPHEREx
}

\def\keywords{%
astronomy --- astrophysics --- journals: individual: JKAS
}

\def\abstracttext{%
The Spectro-Photometer for the History of the Universe, Epoch of Reionization and Ices Explorer (SPHEREx) will provide all-sky spectral survey data covering optical to mid-infrared wavelengths with a spatial resolution of 6\farcs2, which can be widely used to study galaxy formation and evolution. 
We investigate the galaxy-galaxy blending in SPHEREx datasets using the mock galaxy catalogs generated from cosmological simulations and observational data. 
Only $\sim0.7\%$ of the galaxies will be blended with other galaxies in all-sky survey data with a limiting magnitude of 19 AB mag. 
However, the fraction of blended galaxies dramatically increases to $\sim7$--$9\%$ in the deep survey area around the ecliptic poles, where the depth reaches $\sim22$ AB mag. 
We examine the impact of the blending in the number count and luminosity function analyses using the SPHEREx data. 
We find that the number count can be overestimated by up to $10$--$20\%$ in the deep regions due to the flux boosting, suggesting that the impact of galaxy-galaxy blending on the number count is moderate. 
However, galaxy-galaxy blending can marginally change the luminosity function by up to 50\%\ over a wide range of redshifts. 
As we only employ the magnitude limit at $K_s$-band for the source detection, the blending fractions determined in this study should be regarded as lower limits.            
}

\begin{document}
\jkashead 


\section{Introduction}
The Spectro-Photometer for the History of the Universe, Epoch of Reionization and Ices Explorer (SPHEREx) is a future space mission providing unprecedented all-sky spectral survey data covering the optical and near-infrared wavelengths (i.e., 0.75--5\textmu m). 
Owing to the spectral capability of the SPHEREx with a spectral resolution of $\sim40$--150, the survey data will enable us to estimate distances to a large number of galaxies robustly and hence be of great importance in understanding the large-scale structures of the Universe \cite[e.g.][]{Dore2014, Dore2016}. 
In addition, the NIR spectral coverage of the SPHEREx dataset is essential to investigate the stellar properties of galaxies. 
For example, SPHEREx is expected to discover more than thousands of galaxy groups and clusters without additional spectroscopic observation and allow us to measure the stellar masses and star formation rates of individual galaxies. 
Moreover, it will enable us to investigate the luminosity and stellar mass functions of galaxies up to $z\sim0.5$--1. Therefore, SPHEREx will be vital to explore the evolution and formation of galaxies. 

However, due to a relatively large pixel scale and spatial resolution ($\sim6$ arcsec), one needs to take into account galaxy-galaxy blending in investigating the photometric properties of galaxies in the SPHEREx dataset. 
In particular, a Sun-synchronous orbit of the SPHEREx mission will lead to observing areas around the south and north ecliptic poles more frequently, where the photometric data that is $\sim2$--$3$ mag deeper than that in the all-sky area will be available. 
The deep regions with an area of $\sim 200$ deg$^2$ will be more significantly affected by the blending problem. 
This blending can cause bias in determining the large-scale structure and estimating the brightness measurements through flux boosting (see Section \ref{sec:impl}). 
Therefore, it is necessary to quantify this effect with the existing datasets carefully.Note that the sensitivities (5$\sigma$) of SPHEREx data for the point source are around $18.5-19.5$ and $21.5-22$  AB mags in the all-sky and deep regions, respectively \citep{Dore2014, Dore2018}. For simplicity, we assume that the detection limits are 19 and 22 mags in the all-sky survey and deep survey, respectively.

Such a blending issue has been extensively studied in the previous all-sky infrared survey conducted with the Wide-field Infrared Survey Explorer (WISE; \citealp{Wright2010}). 
Angular resolutions of W1 and W2 bands ($\sim$6\farcs1--6\farcs4) are comparable to that of the SPHEREx. 
In addition, its 5$\sigma$ depth ($\sim19.5$ AB mag in W1 band) agrees well with that of the SPHEREx.
From the cross-match between the Galaxy and Mass Assembly (GAMA; \citealp{Driver2009, Driver2011}) survey and WISE dataset, \citet{Cluver2014} demonstrated that the galaxy-galaxy blending (i.e., multiple optical counterparts in a single WISE source) is only severe in the faint end. 
However, this result is based on relatively shallow optical data ($r\leq19.8$ mag) and hence is limited to nearby galaxies ($z<0.5$), which may reveal that the blending fraction of the WISE dataset was underestimated. 
Moreover, because the blending fraction dramatically increases with decreasing galaxy brightness, it needs to be quantified with deep survey data covering high-redshift sources.  

In this paper, we investigate the expected galaxy-galaxy blending fraction in the SPHEREx all-sky survey and deep regions dataset, using both numerical and observational data.
In Section 2, we describe the basic properties of various galaxy catalogs used in this study and our methods to compute the blending fraction. 
The blending fractions from the various datasets are presented in Section 3. 
From this result, we discuss its implication for the statistical properties of galaxies in Section 4. 
We summarize the results in Section 5. 
Throughout this study, we adopt the following cosmological parameters: $H_0=100h=67.4$ km ${\rm s}^{-1}$ ${\rm Mpc}^{-1}$, $\Omega_m=0.315$, and $\Omega_\Lambda=0.685$ \citep{Planck2018}.
All magnitudes are in AB system.

\section{Data and Method}
\subsection{Cosmological Simulation}\label{sec:simdata}
We utilize a mock galaxy catalog from a cosmological simulation to investigate the blending fraction in the SPHEREx survey dataset. 
We specifically adopt the mock catalog to forecast the ultra-wide field of Nancy Grace Roman Space Telescope \citep[Roman,][]{spergel2015}\footnote{\url{https://flathub.flatironinstitute.org/group/sam-forecasts}}, which was produced using 164 snapshots ranging from $z = 0$ to $z = 10$ covering a total area of $\sim2$ deg$^2$, containing at least $25$ million subhalos. 
The simulated galaxies cover a wide range of UV luminosities from $-16$ to $-26$ mag (\citealp{Somerville2021,Yung2022b}).
The halo distribution was extracted from the SmallMultiDarkPlanck (SMDPL) from MultiDark N-body simulation suite. 
SMDPL specifically conducted in a box size of $400 h^{-1}$ Mpc with a particle mass of $9.6 \times 10^{7} h^{-1} M_{\odot}$ (\citealp{Klypin2016}).
{\tt ROCKSTAR} code was used to identify dark matter halos and subhalos complete down to a halo mass of $\sim 10^{10} M_{\odot}$, which contains $\sim 100$ dark matter particles \citep{Behroozi2013}.
Finally, Santa Cruz Semi-Analytic Model (SC-SAM) was adopted to trace the baryonic properties of the galaxies in the halos \citep{Somerville2015}. 
These data work reasonably well in reproducing the luminosity function and number counts up to 25-26 mag derived from the Cosmic Assembly Near-infrared Deep Extragalactic Legacy Survey \citep[CANDELS,][]{Grogin2011}, indicating that it is suitable for our purpose \citep{Somerville2021}.
The lightcone data is available in the form of five distinct datasets, each representing different realizations, and we can employ these datasets for estimating cosmic variance.

\begin{figure}
    \centering
    \includegraphics[width=\linewidth]{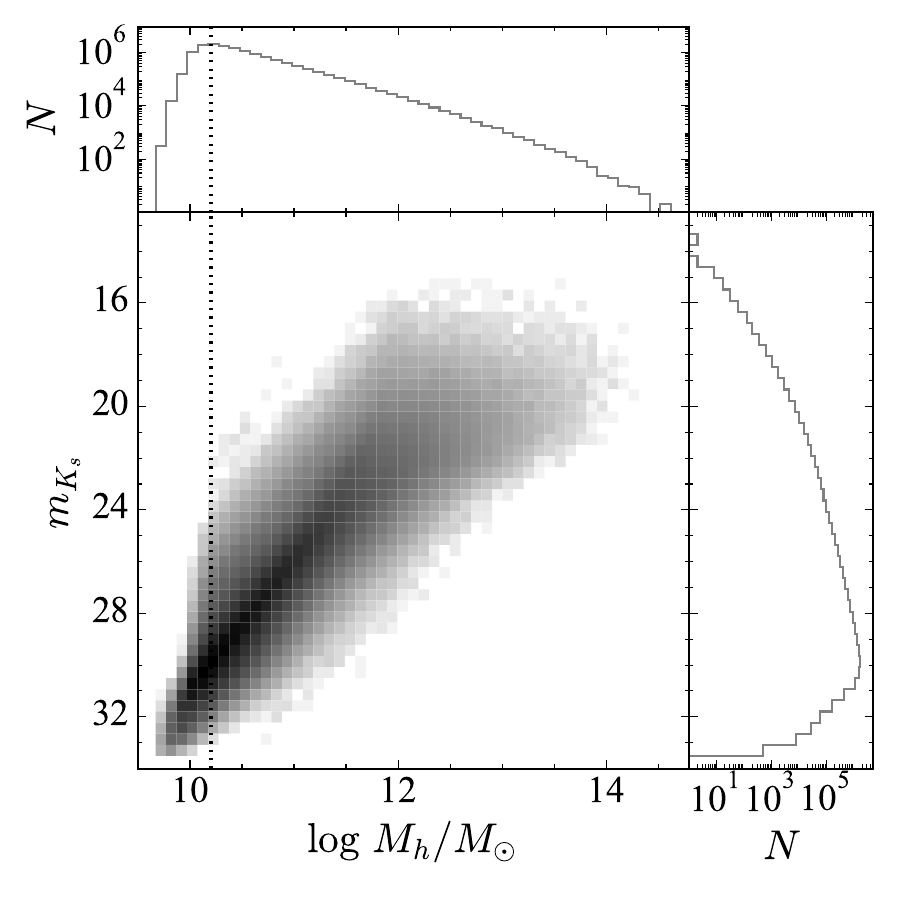}
    \caption{Histograms of halo mass and $K_s$ apparent magnitude of the galaxies in the Santa Cruz Semi-Analytic Model (SC-SAM) ultra-wide field mock catalog. 
    The vertical dotted line denotes the minimum reliable halo mass ($2.2\times10^{10}\,M_\odot$) imposed by the mass resolution limit ($2.2\times10^8\,M_\odot$) of the simulation.}
    \label{fig:jwst-Ks-Mh_2dhist}
\end{figure}

Figure \ref{fig:jwst-Ks-Mh_2dhist} shows the histograms of halo mass and Visible and Infrared Survey Telescope for Astronomy \citep[VISTA,][]{Emerson2004} $K_s$ apparent magnitude (with dust attenuation considered) of galaxies in one ultra-wide lightcone.
The vertical dotted line delineates the mass resolution limit of the simulation, which we take to construct a sample for blending fraction measurements.
$K_s$ apparent magnitudes of galaxies more massive than this mass limit go down to $\sim 33$ mag, much deeper than the detection limit of the deep survey.
However, we further limit the sample with $m_{K_s}<24.8$ to match the magnitude cut of the COSMOS sample, which results in 643,885 galaxies (see Section \ref{sec:obsdata}). 
We also prepare samples of $m_{K_s}<25.8$ (1,157,476 galaxies) and $<26.8$ (2,006,859 galaxies) to confirm if the completeness of the sample of $m_{K_s}<24.8$ is enough to investigate the impact of flux boosting (see Section \ref{sec:impl}).

We note that these data stand as the exclusive resource for investigating the blending issue in the SPHEREx surveys from multiple points of view. 
Firstly, both the mass resolution and box size of the simulation are optimal to emulate the SPHEREx deep surveys.
More importantly, the SC-SAM model continues to trace galaxies even beyond the point where their own halos are no longer identified in simulations.
Such galaxies that have lost their own halos, whether due to physical processes or numerical limitations, are termed orphan galaxies.
In the context of (semi-)analytical modeling of galaxies within halos in $N$-body simulations, it becomes inherently challenging to track the evolution of orphan galaxies.
While a common approach is to cease tracking orphan galaxies and place them at the center of their new host halos, the SC-SAM model stands out by predicting the trajectories of orphan galaxies based on their paths before losing their original host halos \citep{Somerville2008}.
This prevents the situation where higher blending could occur simply due to orphan galaxies being placed at the centers of their new host halos.

Further refinements have been made to accurately replicate the observed clustering on small scales, a pivotal aspect in addressing blending phenomena.
These adjustments involve aligning the distributions of satellite galaxies, including orphan galaxies, with a Navarro-Frank-White (NFW) model \citep{NFW}.
Consequently, the angular two-point correlation functions derived from these modified datasets exhibit good agreement with observations over a broad redshift range, extended to scales as small as $\sim 6$ arcsec (see \citealp{Yung2022} for additional details).

\subsection{Deep Survey Data}\label{sec:obsdata}
\begin{figure}
    \centering
    \includegraphics[width=\linewidth]{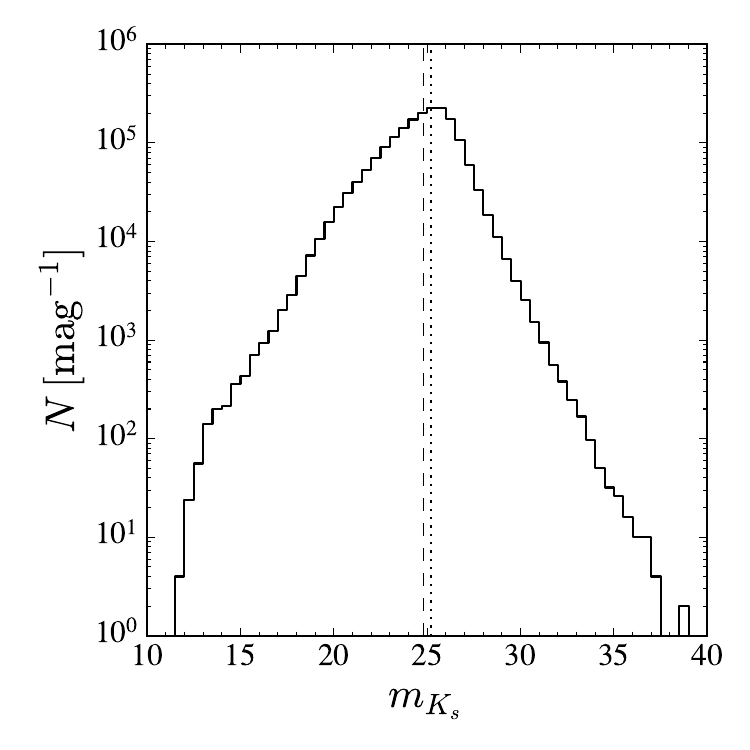}
    \caption{$K_s$ apparent magnitude distribution of objects in the COSMOS2020 catalog within the contamination-free area of 1.27 deg$^2$.
    The vertical dashed and dotted lines denote the limiting magnitudes of UltraVISTA DR4 deep and ultradeep surveys, i.e., $m_{K_s}=24.8$ and $m_{K_s}=25.2$.}
    \label{fig:COSMOS-Ks_hist}
\end{figure}

The Cosmic Evolution Survey \citep[COSMOS,][]{Scoville_2007} has emerged as a fundamental component of extragalactic astronomy, providing multiwavelength photometry for more than one million sources across a 2 deg$^2$ field. 
While multiple public releases of source catalogs have been made available, we use the latest one (COSMOS2020) released in 2022 \citep{COSMOS2020}. 
The main achievements in COSMOS2020 are the substantially deeper optical and near-infrared images obtained from the ongoing Subaru-HSC \citep{Aihara2019} and VISTA-VIRCAM \citep{McCracken2012,Moneti2023} surveys, the reprocessed Spitzer data \citep{Moneti2022}, and the improved astrometry with Gaia \citep{Gaia2016,Gaia2018}. 
Thanks to the deep images, the number of detected sources has largely increased compared to the previous 2015 catalog \citep{Laigle2016} from about half a million to 1.7 million objects across.
Especially the depth in the UltraVISTA DR4, which includes four bands from $Y$ to $K_s$, overlapping with the SPHEREx wavelength coverage, has increased by approximately one magnitude. 
Photometric redshifts are computed for all objects with subpercent and 5\% precision for bright and faint objects, respectively.

Figure \ref{fig:COSMOS-Ks_hist} shows the $K_s$-band magnitude distribution of galaxies that are not contaminated by bright stars or artifacts (selected with \texttt{lp\_type}=0 and \texttt{FLAG\_COMBINED}=0) in the COSMOS field, indicating that the data is complete up to $m_{K_s}\approx25.2$, which corresponds to the $3\sigma$ depth of the ultradeep survey as indicated in Table 1 of \citet{COSMOS2020}. 
However, the ultradeep survey comprises four distinct stripes, whereas the deep survey covers one contiguous area. 
To construct a complete and homogeneous sample, we adopt the magnitude cut of $m_{K_s}<24.8$, the magnitude limit of the deep survey, and confirm that the sample with this magnitude cut does not exhibit any inhomogeneities across the field.
It is important to highlight that the COSMOS data stands out as the sole publicly accessible dataset reaching such a depth. 
For example, while the United Kingdom InfraRed Telescope (UKIRT) Infrared Deep Sky Survey (UKIDSS) Ultra-Deep Survey (UDS) Data Release 11 (DR11, Almaini et al., in preparation) boasts comparable depth among several near-IR survey datasets (with a limiting magnitude of $m_{K_s}\sim25.3$ mag), the non-disclosure of detailed information regarding star/artifact masks and precise survey area measurements renders it unusable for this study.
Additionally, the redshift information in the COSMOS data offers another advantage over alternative survey datasets, facilitating various investigations into blending phenomena, such as the determination of physical associations among blending pairs.
Please refer to \citet{COSMOS2020} and the documents provided along with the catalog\footnote{\url{https://cosmos2020.calet.org}} for more details regarding source classifications, flags, and limiting magnitudes. 

By applying this magnitude cut as well as the flags for galaxies and contamination-free, we are left with 303,275 galaxies in the contamination-free area of 1.27 deg$^2$.
Based on this selection, we estimate blending fractions incurred due to the SPHEREx spatial resolution and further investigate their impact on luminosity functions and number counts in the SPHEREx datasets.

\subsection{Comparison between simulation and survey data}\label{sec:data_compared}

\begin{figure}[h]
    \centering
    \includegraphics[width=\linewidth]{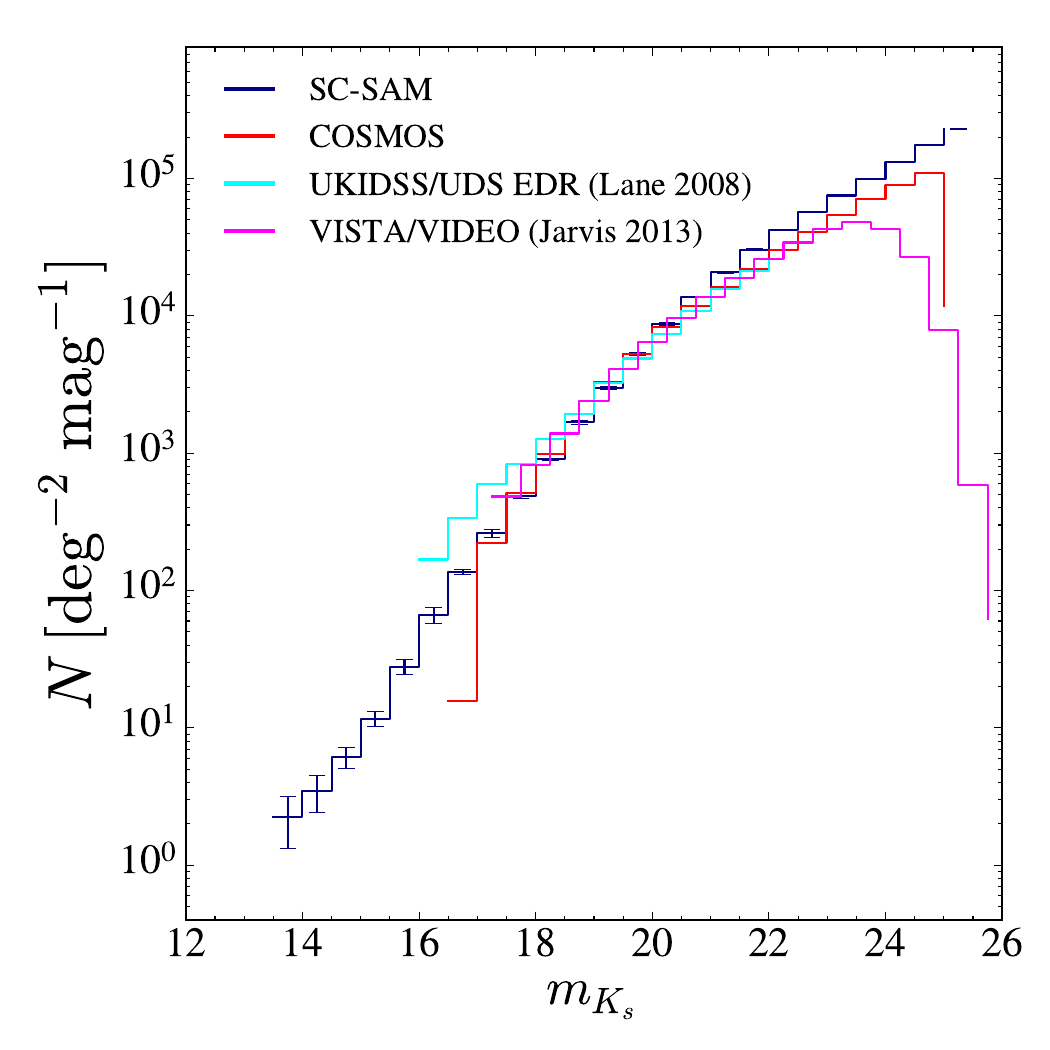}
    \caption{Number of galaxies per deg$^2$ as a function of $K_s$ apparent magnitude of the COSMOS (red) and SC-SAM (blue) samples.
    The results from the United Kingdom InfraRed Telescope (UKIRT) Infrared Deep Sky Survey (UKIDSS) Ultra-Deep Survey (UDS) Early Data Release (EDR, \citealp{Lane2007}) and the Visible and Infrared Survey Telescope for Astronomy (VISTA) Deep Extragalactic Observations (VIDEO, \citealp{Jarvis2013}) are shown as well.
    The errors for the SC-SAM samples are estimated using five different lightcones. 
    The discrepancy between simulation and observation at faint magnitudes could be attributed to either the incompleteness of the observational data or overly populated dark matter halos in simulation.}
    \label{fig:Ks_hist}
\end{figure}

Figure \ref{fig:Ks_hist} shows the comparison between the number counts of the SC-SAM and COSMOS samples constructed for the study (see Sections \ref{sec:simdata} \& \ref{sec:obsdata}) as a function of $K_s$-band magnitude.
The results from the UKIDSS UDS Early Data Release (\citealp{Lane2007}) and the Visible and Infrared Survey Telescope for Astronomy (VISTA) Deep Extragalactic Observations (VIDEO, \citealp{Jarvis2013}) are shown as well.
The photometric completeness correction for the observational data is not applied.
The number count of the COSMOS sample appears smaller than the SC-SAM sample at both bright and faint magnitudes; the reduction in the number count the COSMOS sample at bright magnitudes seems partly due to the saturation limit, typically specified as $m_{K_s}\sim14.2$\footnote{\url{https://www.eso.org/rm/api/v1/public/releaseDescriptions/213}}.
While the difference between the two samples could also be attributed to cosmic variance, the integrated count in the SC-SAM sample with $m_{K_s}<22$ is about 20\% larger than that in the COSMOS sample (50,200 deg$^{-2}$ versus 42,300 deg$^{-2}$; see also \citealp{Somerville2021}).
It is also reported that the clustering in the SC-SAM sample appears stronger compared to observations at scales below 6 arcsec (see Figure 9 of \citealp{Yung2022}).
This may explain the larger blending fractions for the SC-SAM than those in the COSMOS.
However, discerning which, simulation or observation, represents the ground truth is not always evident, particularly close to the sensitivity or resolution limits of the datasets.
Moreover, elucidating the reason for the discrepancy is beyond the scope of this paper.
Instead, we emphasize the importance of leveraging different datasets to reach a generally acceptable conclusion, considering the influence of diverse factors and limitations.
It is also worth noting that the estimated blending fractions from these two samples are considered to represent lower limits for the actual blending fraction. 
This is due to our calculations not incorporating galaxy sizes, which would contribute to further instances of blending.

\subsection{Measurements of Blending Fraction}
We calculate the blending fraction based on the pixel basis. We first construct a grid with a 6.2-arcsec spacing in both R.A. and Decl. directions, which mimic the pixel elements of SPHEREx of the same size \citep{Dore2016}. 
We subsequently count pixels containing more than one galaxy and a blending fraction is defined as the number fraction of such pixels among all pixels associated with galaxies. 
Because we do not know the exact location of the grid, we iterate this procedure 100 times using different grids, whose locations are randomly shifted by a sub-pixel amount each time. 
Then, we take the mean and standard deviation of blending fractions of these 100 iterations to estimate a blending fraction and its uncertainty.
In the case of the SC-SAM sample, we can also use the five lightcones for the error estimate. 
In this experiment, the galaxy is assumed to be a point source, and stars are not considered, which can lead to underestimating blending fractions. In addition, the source detection is determined solely based on the $K_s$ magnitude for simplicity, at which the spectral energy distribution (i.e., $F_\nu$) peaks for galaxies with moderate redshift ($\sim0.5-1.0$; \citealp{Sawicki_2002}). 
Therefore, our estimates will serve only as lower limits of actual values.

\section{Result}\label{sec:res} 
The confusion limit refers to a flux limit at which the cumulative number density of sources, $N(<m_{K_s})$, multiplied by the beam size (i.e., the spatial resolution) is unity \citep{Casey2014}. 
Based on the SC-SAM number counts in Figure 3 and the SPHEREx spatial resolution, this confusion limit is $m_{K_s}\approx23.5$. 
This indicates that a SPHEREx pixel contains, on average, fewer than a single galaxy of $m_{K_s}<23.5$, making blending among galaxies that are brighter than the SPHEREx survey limits (both the all-sky and deep surveys) potentially insignificant. 
However, it is important to note that galaxy clustering could still lead to notable blending. 
Previous studies \citep[e.g.,][]{Patanchon2009,Roseboom2010} that investigated the effect of galaxy clustering on blending generally concluded that the impact of clustering on blending is negligible. 
However, these studies often made simplifying assumptions such as linear clustering that is independent of flux or spectral energy distribution \citep[e.g.,][]{FC2008}. 
In our study, however, the effect of galaxy clustering can be investigated more in-depth by using both high-resolution cosmological simulation data and observational data.

It is worth mentioning that the blending that includes galaxies that are fainter than the survey limits is not discussed in this section.
The SPHEREx pipeline will perform forced photometry by using positional prior information on the measurable SPHEREx sources to compensate for the poor spatial resolution.
Therefore, the main focus is on blending among sources brighter than the survey limits, those compiled in the SPHEREx's reference catalog.
A more comprehensive discussion of blending, including fainter sources, will be presented in Section \ref{sec:impl}.
\subsection{Blending in SPHEREx All-Sky Survey}\label{sec:bf_allsky}
\begin{figure}
    \centering
    \includegraphics[width=\linewidth]{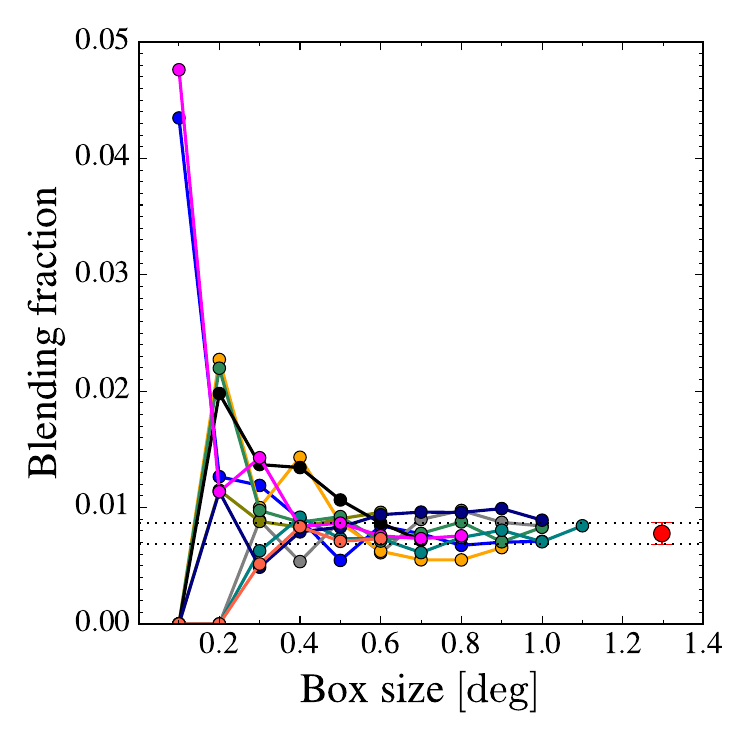}
    \caption{Blending fraction as a function of a square-shaped survey area obtained from the SC-SAM sample.
    Different colors are for different locations of each square in the sky.
    Depending on whether it is located in a crowded or empty region, the blending fraction could vary significantly when the survey area is small.
    As the survey area increases, the blending fraction converges to a value, represented by the red circle with an error bar calculated using the full sample  (horizontal dotted lines).
    The error bar is calculated from 100 different grids.}
    \label{fig:bf_sim_19}
\end{figure}

\begin{figure*}
    \centering
    \includegraphics[width=\textwidth]{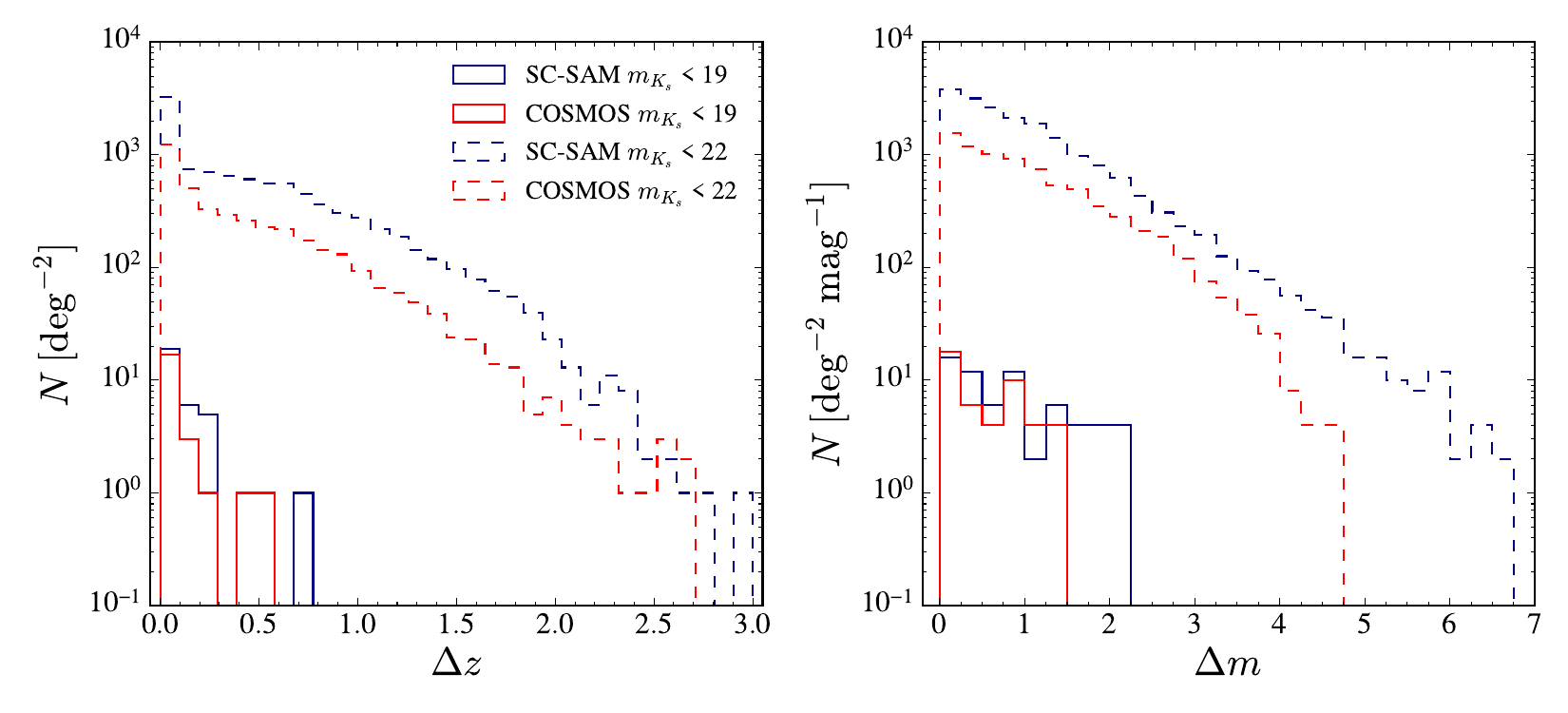}
    \caption{Distributions of redshift difference (left) and $K_s$ apparent magnitude difference (right) of blending pairs found in the SC-SAM (blue) and COSMOS (red) samples of $m_{K_s}<19$ (solid) and $m_{K_s}<22$ (dashed).
    Blending tends to happen among galaxies at similar redshifts and magnitudes.}
    \label{fig:blending-dzdm}
\end{figure*}

We investigate the blending among the galaxies in the SC-SAM sample that are brighter than the limiting magnitude of the SPHEREx all-sky survey, i.e., $m_{K_s}<19$.
It is clear that this neglects blending instances that involve galaxies fainter than the limiting magnitude.
Consequently, the estimated blending fraction might once again fall below the true value.
Nevertheless, this estimate could be employed as a reference to evaluate, for instance, the completeness of the (photometric) redshift catalog of the SPHEREx all-sky survey.

Figure \ref{fig:bf_sim_19} shows how the blending fraction evolves as a function of the survey area.
The survey area is illustrated as a square, initially positioned at an arbitrary location and then enlarged until any side of it aligns with the boundary of the sample.
Lines with different colors correspond to squares positioned at different locations.
When the size of each square is small, the blending fraction within it varies significantly from one to another depending on whether the square contains clusters or voids (i.e., field-to-field variation).
As the survey area increases in size, the blending fractions tend to converge towards a specific value, represented by the red-filled circle with an error bar on the right.
This value is calculated using the full sample, and the error is estimated from 100 different grids.
We further assessed blending fractions using the other four lightcone datasets, as the convergence observed in Figure \ref{fig:bf_sim_19} with increasing survey area may be attributed to a mere increase in overlapping objects.
The dispersion among blending fraction estimates across the five lightcone datasets appears minimal (0.11\%), akin to that observed with 100 grids (0.09\%).
This suggests that despite the limited survey area covered by the lightcone or the COSMOS data, they are sufficiently expansive to derive a representative blending fraction estimate.
Consequently, we opted to exclude consideration of different lightcone mock realizations and instead analyze scatter from varying grid realizations in subsequent analyses.

The blending among galaxies of $m_{K_s}<19$ over the all-sky would not be different from the estimate with the full sample; it is expected to be as low as 1\%.
Similarly, we obtain a low blending fraction with COSMOS galaxies of $m_{K_s}<19$, consistent with the result from the SC-SAM sample.
There are no cases of blending of three or more galaxies, mainly due to the small number statistics. 

We further examine the differences in redshifts and apparent magnitudes of blending pairs, which are denoted by solid lines in the left and right panels of Figure \ref{fig:blending-dzdm}, respectively.
The distributions indicate that the galaxies in a blending pair are likely to have similar redshifts and apparent magnitudes, although the trend in the magnitude difference that prefers small differences is not as strong as that in the redshift difference.
This suggests that a significant portion of blending pairs are genuine pairs with similar masses that are physically associated.
This implies that the blending issue will be critical for the study of galaxy groups/clusters, which will be one of our future works.

\subsection{Blending in SPHEREx Deep Survey}\label{sec:bf_deep}
We repeat the analyses in the previous Section but with the sample of $m_{K_s}<22$ (i.e., applying the depth of the deep survey). 
The blending fractions in the deep survey are increased significantly to $\sim10\%$ from $\sim1\%$ (the blending fractions estimated for the all-sky survey) as summarized in Table \ref{tab:bf}.
About $\sim 9\%$ of the blended pixels contain three or more galaxies. 
As expected from the comparison between the SC-SAM and COSMOS samples in Section \ref{sec:data_compared}, the blending fraction in the SC-SAM sample is higher than that estimated in the COSMOS sample (9.6\% versus 7\%). 
Both of the two estimates suggest that the blending fraction in the deep regions will be non-negligible, unlike in the all-sky survey, and thus should be taken into account for analyses such as number counts and luminosity functions (see Section \ref{sec:impl}).

We also examine the redshift/magnitude differences between blending pairs (dashed lines in Figure \ref{fig:blending-dzdm}), which reinforce the conclusion drawn from the blending pairs in the all-sky survey; blending tends to occur between galaxies of similar masses that are physically associated.  
It is worth noting that the distribution of redshift difference constructed from the SC-SAM sample is significantly more peaked at small values than that from the COSMOS sample.
This might indicate that galaxies in groups at high redshifts, which are separated by small angular separations, are not entirely sampled in the COSMOS sample.
Galaxies at high redshifts, which may be blocked by larger foreground galaxies, could also be absent from the COSMOS sample. This leads to fewer blending instances with large redshift differences compared to the SC-SAM sample.
This may contribute to the incompleteness of the COSMOS sample and thus lead to its smaller sample size (than the SC-SAM sample) when a faint limiting magnitude is considered (see Section \ref{sec:data_compared}).

\begin{table}[t!]
\caption{Blending fraction estimates for SPHEREx surveys\label{tab:bf}}
\centering
\begin{tabular}{lrr}
\toprule
Data    & $m_{K_s}<19$ & $m_{K_s}<22$ \\
\midrule
COSMOS      & 0.73  $\pm$ 0.11 \% & 7.06 $\pm$ 0.10 \%   \\
SC-SAM      & 0.77 $\pm$ 0.09 \% & 9.13 $\pm$ 0.06\%   \\
\bottomrule
\end{tabular}
\tabnote{Blending fractions estimated using the SC-SAM and COSMOS samples that are limited by the flux limits of the SPHEREx all-sky and deep surveys.
The errors are estimated using 100 realizations of the SPHEREx pixel grid.
}
\end{table}

\section{Implication}\label{sec:impl}
The blending is expected to increase the detectable sources, which may arise due to flux boosting caused by multiple sources below the detection limit coincidentally residing in the same pixel, resulting in their total flux surpassing the detection limit.
It is important to highlight that various factors contribute to flux boosting; for example, photometric uncertainties can elevate fluxes and consequently impact the number of sources, especially the number of sources near the detection limit \citep[i.e., Eddington bias,][]{Eddington1913}.
In our investigation, we focus on examining the influence of flux boosting caused by multiple sources (as opposed to that induced by photometric uncertainties) on the source number counts and luminosity functions in the two surveys of SPHEREx.

\begin{figure}[t]
    \centering
    \includegraphics[width=\linewidth]{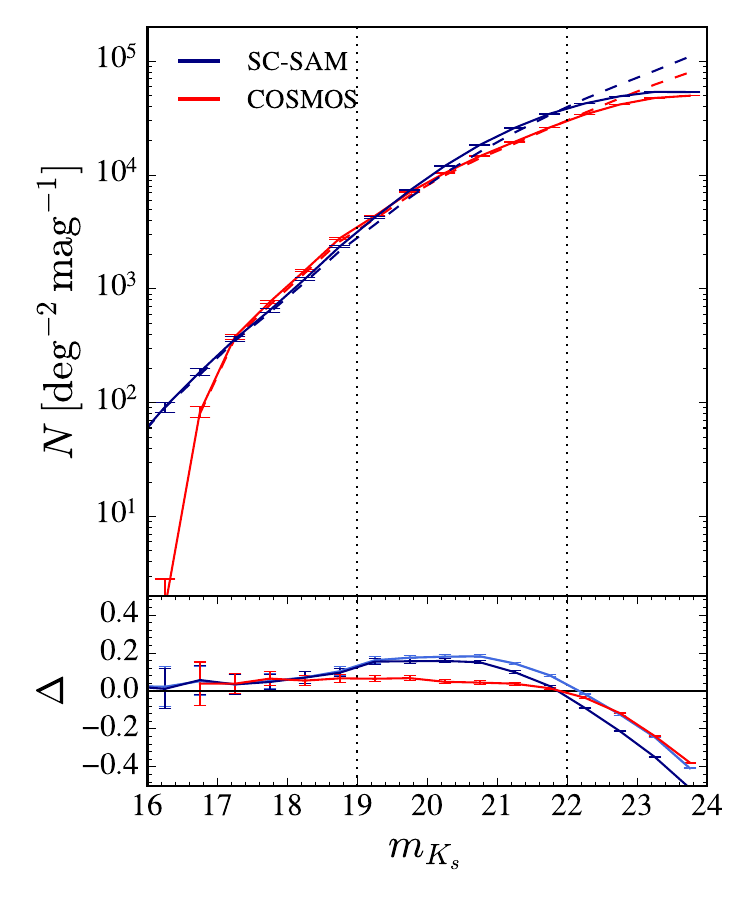}
    \caption{(Top) Source number counts as a function of $K_s$ apparent magnitude obtained from the SC-SAM (blue) and COSMOS (red) samples of $m_{K_s}<24.8$.
    The true number counts ($N_\mathrm{true}$) are denoted by dashed lines, and the number counts with the blending effect ($N_\mathrm{blend}$) are by solid lines.
    Two vertical dotted lines denote the detection limit of SPHEREx all-sky and deep surveys.
    (Bottom) The relative difference between the number counts with the blending effect and the true number counts, defined as $\Delta=(N_\mathrm{blend}-N_\mathrm{true})/N_\mathrm{true}$. 
    Blending results in over-counts at bright magnitudes and under-counts at faint magnitudes.
    The results of the SC-SAM samples of $m_{K_s}<25.8$ and $<26.8$ are presented in lighter blue colors, which overlap almost perfectly.
    While showing a slight difference from the result of the $m_{K_s}<24.8$ sample, the overall trend is not significantly different.
    Errors are calculated based on the bootstrap resampling method.}
    \label{fig:numcount}
\end{figure}

\begin{figure*}[h]
    \centering
    \includegraphics[scale=0.9, width=\linewidth]{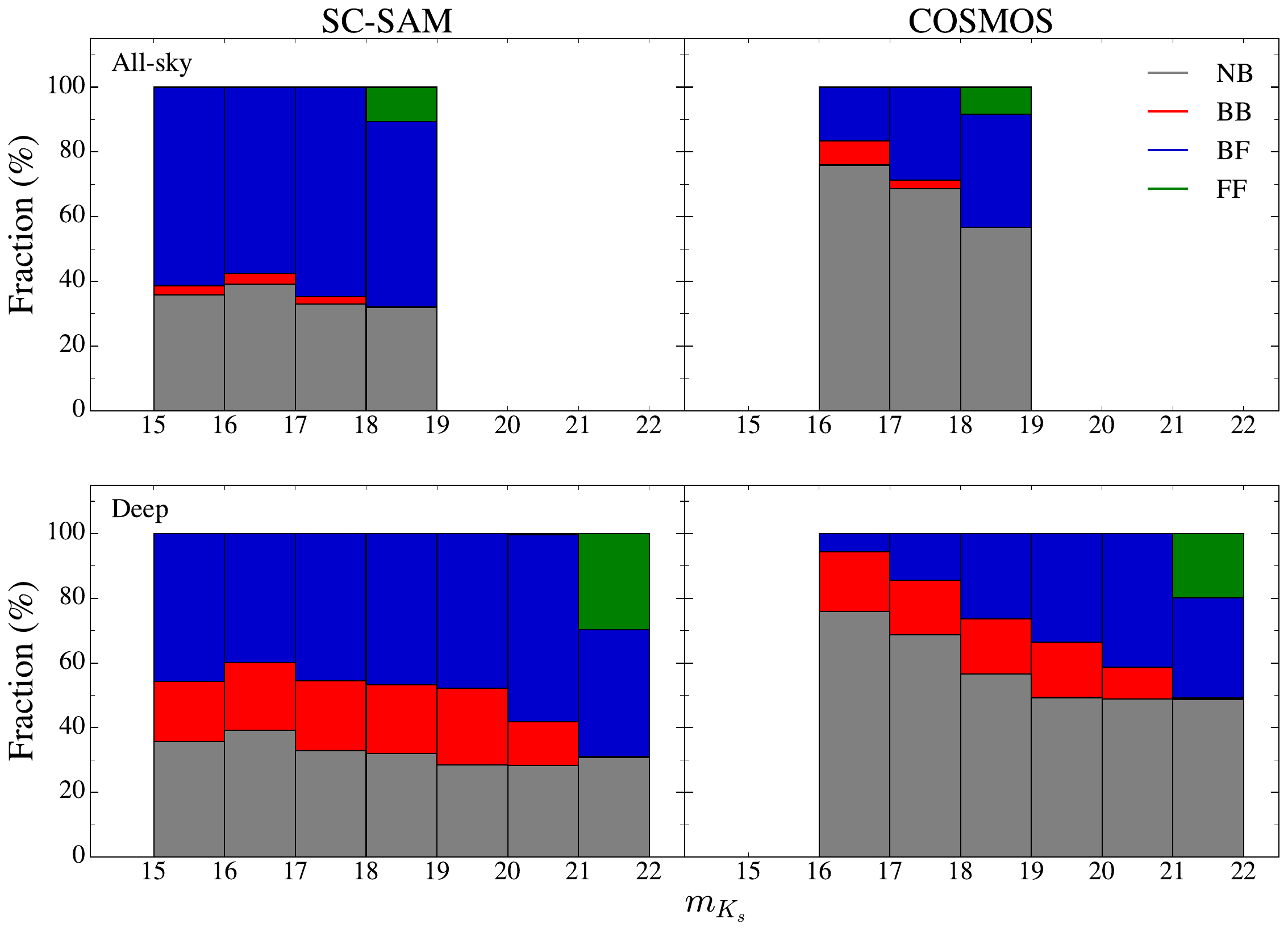}
    \caption{Demography of pixels identified as sources in the all-sky (top) and deep (bottom) surveys in the SC-SAM (left) and COSMOS (right) samples of $m_{K_s}<24.8$. These pixels are categorized into four cases based on the brightness levels of the first and second brightest galaxies: pixels with no blending (NB, gray), blending of bright-bright galaxies (BB, red), blending of bright-faint galaxies (BF, blue), and blending of faint-faint galaxies (FF, green). Here ``bright'' and ``faint'' are defined with respect to each survey limit.
    Blending of bright-bright galaxies is not a major blending event, especially in the all-sky survey.
    In most cases, blending occurs between a bright galaxy and one or several faint galaxies.
    The discrepancies between the SC-SAM and COSMOS samples are somewhat expected, given the discrepancies in the number of faint galaxies and the clustering on small scales in the two samples, as discussed in Section \ref{sec:data_compared}.}
    \label{fig:bf_demography}
\end{figure*}

\subsection{Number Count}\label{sec:numcnt}
To investigate the impact of the flux boosting on the source number count in the SPHEREx surveys, we require catalogs that include objects much fainter than the survey's detection limit (ideally a catalog of all existing objects). 
In practice, we are limited to certain magnitude limits to which the samples are complete (e.g., $K_s<24.8$ in the case of the COSMOS sample). 
Although this limit is almost three magnitudes deeper than the limiting magnitude of the deep survey, we test whether the depth is sufficient for quantifying the flux-boosting effect in the SPHEREx surveys using the two deeper samples prepared from the SC-SAM ultra-wide lightcone data.

The top panel of Figure \ref{fig:numcount} presents the number of sources (i.e., galaxies or SPHEREx pixels) as a function of magnitude. The blue and red dashed lines are the number of {\it galaxies} ($N_\mathrm{true}$) from the SC-SAM and COSMOS samples, respectively (i.e., true source counts that are the same as Figure \ref{fig:Ks_hist}).
The solid lines represent the count of {\it pixels} when accounting for the blending effects ($N_\mathrm{blend}$). When blending occurs, we compute
the total flux within each pixel that contains multiple galaxies. 
The bottom panel displays the relative difference between the solid and dashed lines, i.e., $\Delta=(N_\mathrm{blend}-N_\mathrm{true})/N_\mathrm{true}$, of the corresponding color. Errors are calculated based on the bootstrap resampling method.
Although the results from the SC-SAM and COSMOS samples exhibit discrepancies due to the incompleteness of either dataset, a consistent conclusion regarding the influence of blending on number counts can be reached: source count boosting appears dominant at bright magnitudes (e.g., $m_{K_s}\lesssim22$), whereas source count reduction dominates at faint magnitudes (e.g., $m_{K_s}\gtrsim22$).
The source count reduction at faint magnitudes is due to the fact that blending tends to occur within a group/cluster and thus faint galaxies in a group/cluster will not be counted as individual sources but rather as a single source with a brighter magnitude.
While the all-sky survey would end up with more sources by $\lesssim 10\%$ in each magnitude bin, such artificial excess in source count in the deep survey could extend to $\sim 20\%$ at faint magnitudes.

\begin{figure*}[h]
    \centering
    \includegraphics[width=\textwidth]{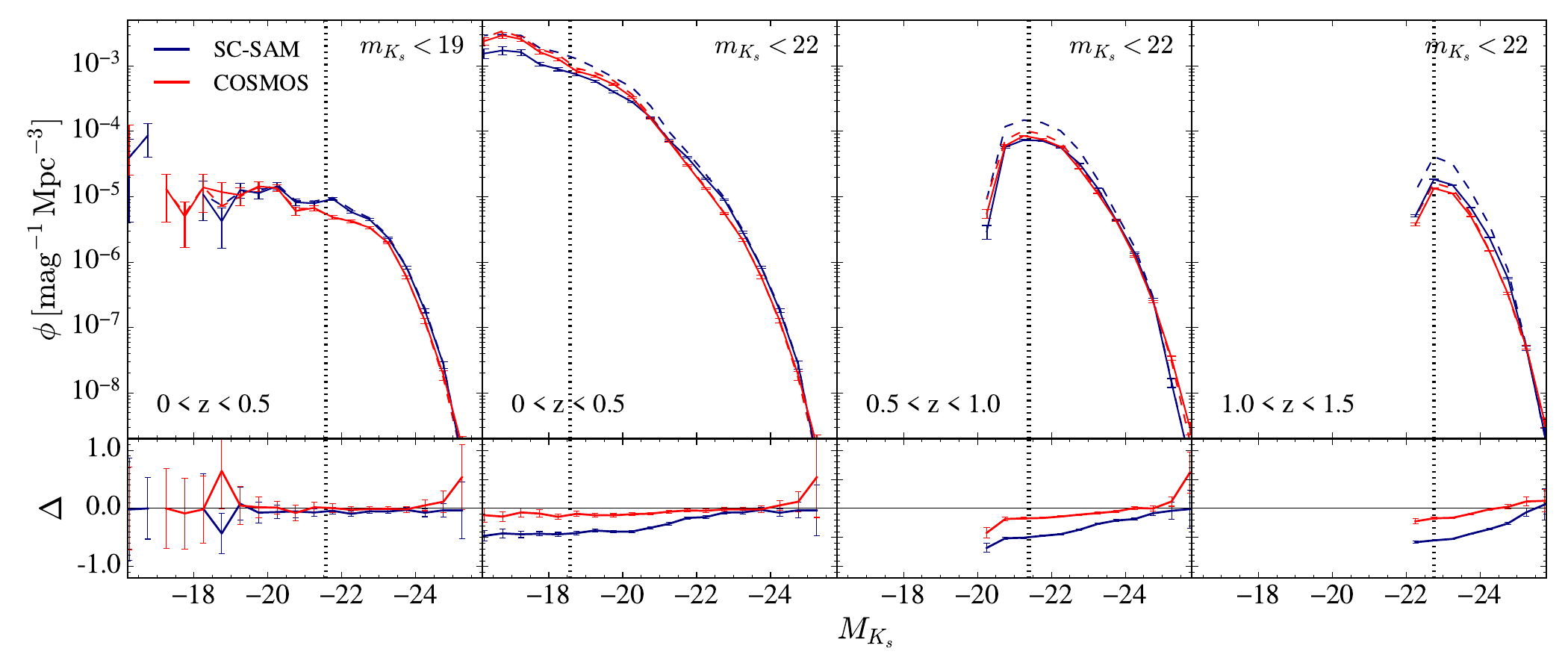}
    \caption{Binned luminosity functions with (solid) and without (dashed) the effect of blending at various redshifts.
    The leftmost panel displays the luminosity function derived from the all-sky survey at the lowest redshift bin.
    The remaining panels show the luminosity functions for the deep survey at three redshift bins. 
    The relative difference ($\Delta=(\phi_\mathrm{blend}-\phi_\mathrm{true})/\phi_\mathrm{true}$) between the luminosity functions with ($\phi_\mathrm{blend}$) and without ($\phi_\mathrm{true}$) the effect of blending is presented in the bottom panel of each main panel.
    Vertical dotted lines represent the limiting absolute magnitudes that correspond to the flux limits of the deep and all-sky surveys, respectively.
    The conversion between apparent and absolute magnitudes is based on the central redshift in each bin. 
    Blending has minimal ($<1\sigma$) effects on the luminosity function of the all-sky survey $m_{K_s}<19$, but it results in non-negligible ($>1\sigma$) underestimates in the luminosity functions obtained from the deep surveys $m_{K_s}<22$ across all magnitudes.
    This effect becomes more pronounced at higher redshifts.}
    \label{fig:LF}
\end{figure*}

Two lines in lighter blue colors in the bottom panel, which are overlapping almost perfectly and thus indistinguishable amongst themselves, correspond to the deeper SC-SAM samples of $m_{K_s}<25.8$ and $m_{K_s}<26.8$.
These two lines are additionally presented to assess the validity of the $m_{K_s}<24.8$ sample for this study, as mentioned earlier.
The difference between the blue (the $m_{K_s}<24.8$ sample) and lighter blue (the $m_{K_s}<25.8$ and $m_{K_s}<26.8$ samples) lines appears clear, showing the actual transition from overestimate to underestimate occurs at a magnitude fainter (by $\sim$0.4 mag) than the magnitude derived from the $m_{K_s}<24.8$ sample.
This indicates that while the $m_{K_s}<24.8$ sample seems acceptable to investigate the impact of blending on the source number count of the all-sky survey, a deeper sample such as that of $m_{K_s}<25.8$ at least is required for the deep survey. 
However, given that the overall trend remains consistent across all three samples, the conclusion that the non-negligible excess in the source number count is expected in both all- and deep-sky surveys is still valid.

To compare with the blending fractions estimated in Section \ref{sec:res}, where faint sources are not considered, we assess blending fractions using the SC-SAM and the COSMOS samples of $m_{K_s}<24.8$.
Although the blending fraction is defined in the same way as previously described, i.e., the fraction of pixels featuring blending, the distinction lies in the application of the survey flux limits to pixels instead of galaxies.
Consequently, galaxies with fainter fluxes may contribute to the total flux of a pixel.
The blending fractions exhibit a significant increase by considering faint sources, from  0.77 and 9.13\% (refer to Table \ref{tab:bf}) to 67.4 and 70.1\% for the all-sky and deep surveys, respectively, for the SC-SAM sample.
In the case of the COSMOS sample, it increases from 0.73 and 7.06\% to 40.5 and 50.5\%.
It is primarily due to the large number of faint sources.

Figure \ref{fig:bf_demography} shows the demography of pixels identified as sources in the all-sky (top) and deep (bottom) surveys in the SC-SAM (left) and COSMOS (right) samples, respectively, in each $K_s$-band magnitude bin.
We categorize these pixels into four cases based on the brightness levels of the first and second brightest galaxies: pixels with no blending (NB, gray), blending of bright-bright galaxies (BB, red), blending of bright-faint galaxies (BF, blue), and blending of faint-faint galaxies (FF, green).
Here, ``bright'' and ``faint'' are defined with respect to each survey limit, as has been the case thus far.
As seen in Section \ref{sec:res}, the blending of bright-bright galaxies is not prevalent in the all-sky survey but becomes significant in the deep survey.
In most instances, blending occurs between a bright galaxy and one or several faint galaxies, as indicated by the size of the blue bar compared to the red and green ones.
These blending instances are not identified as blending in Section \ref{sec:res}.
Although their impact on photometry will be minimal at bright magnitudes, it becomes noteworthy near faint magnitudes close to the survey limits.
As noted earlier, because of the discrepancy in the number of faint sources and the clustering on small scales in the SC-SAM and COSMOS samples (see Section \ref{sec:data_compared}), the blending fractions in the two samples with the fainter magnitude cut exhibit larger differences than those in Section \ref{sec:res}.
Cases where only sources fainter than the detection limits contribute to the blending, their total flux exceeding the survey limits and consequently falsely identified as a source, account for 7.7 and 16.8\% in the all-sky and deep surveys, respectively, in the SC-SAM sample, and 6.5 and 10.4\% in the COSMOS sample. However, its impact may be minimal in the SPHEREx survey, as the photometric measurements will be only conducted for the known sources without new source detection using the SPHEREx imaging data.

\subsection{Luminosity Function}

We further delve into the influence of blending on the estimation of the luminosity functions at different redshifts. 
Similar to the earlier section's analysis of number counts, we construct binned luminosity functions with and without blending effects.  
This is accomplished using the SC-SAM and COSMOS samples of $m_{K_s}<24.8$. 
When blending is considered, we compute the total flux within each pixel that contains multiple galaxies as done in Section \ref{sec:numcnt}.
If the magnitude difference between the brightest and second brightest galaxies in a blended pixel is greater than 2.5 mag (i.e., a brightness difference of 10 times), we assign the redshift of the brightest galaxy to that pixel.
Pixels with smaller magnitude differences are excluded from the luminosity function construction as relatively large redshift uncertainties are expected in the real observations.
We also checked other thresholds of the magnitude difference for accurate redshift measurements (i.e., $\Delta m_{\rm thres}=$0.5, 1, and 2 mag), which will be discussed at the end of this section. 

We show the resulting luminosity functions together with their relative differences in Figure \ref{fig:LF}; the leftmost panel shows the luminosity function constructed from the all-sky survey at $0<z<0.5$ and the other panels show those from the deep survey across three redshift bins ranging from 0 to 1.5. 
The identical color scheme and line styles used in Figure \ref{fig:numcount} are applied.
Vertical dotted lines represent the limiting absolute magnitudes corresponding to each survey's flux limit, below which the luminosity functions are underestimated due to the incompleteness of the data.
As the aim of this analysis is not to robustly construct luminosity functions beyond the survey limits but rather to assess the direct impact of blending on them, we refrain from employing the $1/V_\mathrm{max}$ technique \citep[e.g.,][]{Felten1977}.
Instead, we present the binned luminosity functions in their original form.
The conversion between apparent and absolute magnitudes of the survey limits is based on the central redshift in each bin.
While the results of the SC-SAM and COSMOS samples do not completely agree with each other, we will prioritize the result of the SC-SAM sample as it is unaffected by the redshift uncertainties, unlike the COSMOS sample.

Contrary to the findings in the number count analysis, blending mostly leads to underestimates in luminosity functions. 
These underestimates are minimal in the luminosity function of the all-sky survey, but become more significant in the deep survey, reaching about 50\% at low luminosities.
At a given luminosity, the underestimate increases as it goes to higher redshifts.
The overestimates observed in relatively bright magnitudes in the number count analysis do not manifest significantly in the luminosity functions.
This can be attributed to the exclusion of pixels when the blending occurs between the galaxies with similar brightnesses (i.e., $\Delta m \leq 2.5$ mag), accounting for a significant portion of blended pixels.

Given that blending primarily occurs between physically associated galaxies (e.g., see Figure \ref{fig:blending-dzdm}), it is more pronounced at higher redshifts where the physical extent per unit angular separation is larger.
Galaxies that are not physically associated also have a chance to be blended, as a result of accidental alignment along the line of sight, although the likelihood of such occurrences is relatively low.
Such blending is likely to involve low-mass galaxies that are an abundant population at any redshift.
A flux-limited survey will sample more low-mass galaxies at lower redshifts, and thus the blending between a lower-mass galaxy at a lower redshift and a higher-mass galaxy at a higher redshift will be a common occurrence in such a scenario.
Such blending will have a larger impact when it is accounted for the higher-redshift luminosity function (unless it is excluded for its small magnitude difference/large redshift uncertainty); it goes for a less-abundant, higher-luminosity population where even minor over/under-estimates can have a significant impact.
These factors can explain the more substantial impact of blending at higher redshifts.

The effect of blending on luminosity functions can change based on the magnitude difference threshold required for accurate redshift estimates. We have verified this change by reconstructing luminosity functions with different $\Delta m_{\rm thres}$ values of 0.5, 1, and 2. If we adopt a smaller $\Delta m_{\rm thres}$ and consequently retain more blended pixels in luminosity functions, we observe overestimations in luminosity functions in the high-luminosity end due to the flux boosting of blended pixels. However, the trend in the low-luminosity end almost remains the same.

\section{Summary and Overlook}
Using the mock galaxy catalog from the cosmological simulation and COSMOS2020 catalog, we find that the galaxy-galaxy blending fraction in the all-sky SPHEREx data is $\sim0.7\%$. As expected, the blending fraction substantially increases up to $\sim7-9\%$ in the SPHEREx deep data, covering the area of 200 deg$^{2}$ around the north and south ecliptic poles. However, these values need to be considered as lower limits of the blending fraction because the galaxy is assumed to be a point source and the source detection is determined only in the $K_S$ band. The blendings are most likely to occur by the galaxies with similar masses and redshifts, suggesting that the blended galaxies are physically associated with each other. This finding suggests that the blending can play an essential role when identifying the members in galaxy groups or clusters.  

The galaxy-galaxy blending can cause increases in the source number counts by up to $20\%$ in the SPHEREx deep data, whereas the impact from the blending in the all-sky data is almost negligible. On the contrary, the blending can marginally change the shape of the luminosity function, in the sense that the LF is underestimated at the low-luminosity end and overestimated at the high-luminosity end. Therefore, this effect should be taken into account when the LF is estimated with the SPHEREx datasets. However, we note that the blending fraction can be underestimated because neither the size of the galaxy nor the blending due to foreground stars are considered in our experiments. Therefore, our estimations on the blending fraction need to be taken as lower limits.      

 As the SPHEREx imaging data will be heavily undersampled, the photometry is applied only to the known objects from the previous data (i.e., forced photometry). Therefore, the impact of the blending may not be as severe as we expected. However, to maximize the photometry accuracy through the deblending, it is essential to know the prior distribution of blended galaxies. Therefore, this study can provide useful insights into this matter. In addition, the bias and uncertainties of the flux measurements attributed to the galaxy-galaxy blending may naturally introduce a non-linear effect on the photo-z estimation, which can be further studied in future work.


\acknowledgments
This work was supported by the National Research Foundation of Korea (NRF), through grants funded by the Korean government (MSIT) (Nos. 2021R1C1C1013580, 2021R1C1C2091550, 2022R1A4A3031306, 2023R1A2C1006261, and RS-2023-00250862). 



\bibliography{ref}

\begin{thebibliography}{}
\expandafter\ifx\csname natexlab\endcsname\relax\def\natexlab#1{#1}\fi
\providecommand{\url}[1]{\href{#1}{#1}}
\providecommand{\dodoi}[1]{doi:~\href{http://doi.org/#1}{\nolinkurl{#1}}}
\providecommand{\doeprint}[1]{\href{http://ascl.net/#1}{\nolinkurl{http://ascl.net/#1}}}
\providecommand{\doarXiv}[1]{\href{https://arxiv.org/abs/#1}{\nolinkurl{https://arxiv.org/abs/#1}}}
\providecommand{\dodoilink}[2]{\href{http://doi.org/#1}{#2}}
\providecommand{\doadslink}[2]{\href{#1}{#2}}

\bibitem[{{Aihara} {et~al.}(2019){Aihara}, {AlSayyad}, {Ando}, {Armstrong},
  {Bosch}, {Egami}, {Furusawa}, {Furusawa}, {Goulding}, {Harikane}, {Hikage},
  {Ho}, {Hsieh}, {Huang}, {Ikeda}, {Imanishi}, {Ito}, {Iwata}, {Jaelani},
  {Kakuma}, {Kawana}, {Kikuta}, {Kobayashi}, {Koike}, {Komiyama}, {Li},
  {Liang}, {Lin}, {Luo}, {Lupton}, {Lust}, {MacArthur}, {Matsuoka}, {Mineo},
  {Miyatake}, {Miyazaki}, {More}, {Murata}, {Namiki}, {Nishizawa}, {Oguri},
  {Okabe}, {Okamoto}, {Okura}, {Ono}, {Onodera}, {Onoue}, {Osato}, {Ouchi},
  {Shibuya}, {Strauss}, {Sugiyama}, {Suto}, {Takada}, {Takagi}, {Takata},
  {Takita}, {Tanaka}, {Terai}, {Toba}, {Uchiyama}, {Utsumi}, {Wang}, {Wang}, \&
  {Yamada}}]{Aihara2019}
{Aihara}, H., {AlSayyad}, Y., {Ando}, M., {et~al.} 2019, \pasj, 71, 114

\bibitem[{{Behroozi} {et~al.}(2013){Behroozi}, {Wechsler}, \&
  {Wu}}]{Behroozi2013}
{Behroozi}, P.~S., {Wechsler}, R.~H., \& {Wu}, H.-Y. 2013, \apj, 762, 109

\bibitem[{{Casey} {et~al.}(2014){Casey}, {Narayanan}, \& {Cooray}}]{Casey2014}
{Casey}, C.~M., {Narayanan}, D., \& {Cooray}, A. 2014, \physrep, 541, 45

\bibitem[{{Cluver} {et~al.}(2014){Cluver}, {Jarrett}, {Hopkins}, {Driver},
  {Liske}, {Gunawardhana}, {Taylor}, {Robotham}, {Alpaslan}, {Baldry}, {Brown},
  {Peacock}, {Popescu}, {Tuffs}, {Bauer}, {Bland-Hawthorn}, {Colless},
  {Holwerda}, {Lara-L{\'o}pez}, {Leschinski}, {L{\'o}pez-S{\'a}nchez},
  {Norberg}, {Owers}, {Wang}, \& {Wilkins}}]{Cluver2014}
{Cluver}, M.~E., {Jarrett}, T.~H., {Hopkins}, A.~M., {et~al.} 2014, \apj, 782,
  90

\bibitem[{{Dor{\'e}} {et~al.}(2014){Dor{\'e}}, {Bock}, {Ashby}, {Capak},
  {Cooray}, {de Putter}, {Eifler}, {Flagey}, {Gong}, {Habib}, {Heitmann},
  {Hirata}, {Jeong}, {Katti}, {Korngut}, {Krause}, {Lee}, {Masters},
  {Mauskopf}, {Melnick}, {Mennesson}, {Nguyen}, {{\"O}berg}, {Pullen},
  {Raccanelli}, {Smith}, {Song}, {Tolls}, {Unwin}, {Venumadhav}, {Viero},
  {Werner}, \& {Zemcov}}]{Dore2014}
{Dor{\'e}}, O., {Bock}, J., {Ashby}, M., {et~al.} 2014, arXiv e-prints,
  arXiv:1412.4872

\bibitem[{{Dor{\'e}} {et~al.}(2016){Dor{\'e}}, {Werner}, {Ashby}, {Banerjee},
  {Battaglia}, {Bauer}, {Benjamin}, {Bleem}, {Bock}, {Boogert}, {Bull},
  {Capak}, {Chang}, {Chiar}, {Cohen}, {Cooray}, {Crill}, {Cushing}, {de
  Putter}, {Driver}, {Eifler}, {Feng}, {Ferraro}, {Finkbeiner}, {Gaudi},
  {Greene}, {Hillenbrand}, {H{\"o}flich}, {Hsiao}, {Huffenberger}, {Jansen},
  {Jeong}, {Joshi}, {Kim}, {Kim}, {Kirkpatrick}, {Korngut}, {Krause}, {Kriek},
  {Leistedt}, {Li}, {Lisse}, {Mauskopf}, {Mechtley}, {Melnick}, {Mohr},
  {Murphy}, {Neben}, {Neufeld}, {Nguyen}, {Pierpaoli}, {Pyo}, {Rhodes},
  {Sandstrom}, {Schaan}, {Schlaufman}, {Silverman}, {Su}, {Stassun}, {Stevens},
  {Strauss}, {Tielens}, {Tsai}, {Tolls}, {Unwin}, {Viero}, {Windhorst}, \&
  {Zemcov}}]{Dore2016}
{Dor{\'e}}, O., {Werner}, M.~W., {Ashby}, M., {et~al.} 2016, arXiv e-prints,
  arXiv:1606.07039

\bibitem[{{Dor{\'e}} {et~al.}(2018){Dor{\'e}}, {Werner}, {Ashby}, {Bleem},
  {Bock}, {Burt}, {Capak}, {Chang}, {Chaves-Montero}, {Chen}, {Civano},
  {Cleeves}, {Cooray}, {Crill}, {Crossfield}, {Cushing}, {de la Torre},
  {DiMatteo}, {Dvory}, {Dvorkin}, {Espaillat}, {Ferraro}, {Finkbeiner},
  {Greene}, {Hewitt}, {Hogg}, {Huffenberger}, {Jun}, {Ilbert}, {Jeong},
  {Johnson}, {Kim}, {Kirkpatrick}, {Kowalski}, {Korngut}, {Li}, {Lisse},
  {MacGregor}, {Mamajek}, {Mauskopf}, {Melnick}, {M{\'e}nard}, {Neyrinck},
  {{\"O}berg}, {Pisani}, {Rocca}, {Salvato}, {Schaan}, {Scoville}, {Song},
  {Stevens}, {Tenneti}, {Teplitz}, {Tolls}, {Unwin}, {Urry}, {Wandelt},
  {Williams}, {Wilner}, {Windhorst}, {Wolk}, {Yorke}, \& {Zemcov}}]{Dore2018}
{Dor{\'e}}, O., {Werner}, M.~W., {Ashby}, M. L.~N., {et~al.} 2018, arXiv
  e-prints, arXiv:1805.05489

\bibitem[{{Driver} {et~al.}(2009){Driver}, {Norberg}, {Baldry}, {Bamford},
  {Hopkins}, {Liske}, {Loveday}, {Peacock}, {Hill}, {Kelvin}, {Robotham},
  {Cross}, {Parkinson}, {Prescott}, {Conselice}, {Dunne}, {Brough}, {Jones},
  {Sharp}, {van Kampen}, {Oliver}, {Roseboom}, {Bland-Hawthorn}, {Croom},
  {Ellis}, {Cameron}, {Cole}, {Frenk}, {Couch}, {Graham}, {Proctor}, {De
  Propris}, {Doyle}, {Edmondson}, {Nichol}, {Thomas}, {Eales}, {Jarvis},
  {Kuijken}, {Lahav}, {Madore}, {Seibert}, {Meyer}, {Staveley-Smith},
  {Phillipps}, {Popescu}, {Sansom}, {Sutherland}, {Tuffs}, \&
  {Warren}}]{Driver2009}
{Driver}, S.~P., {Norberg}, P., {Baldry}, I.~K., {et~al.} 2009, Astronomy and
  Geophysics, 50, 5.12

\bibitem[{{Driver} {et~al.}(2011){Driver}, {Hill}, {Kelvin}, {Robotham},
  {Liske}, {Norberg}, {Baldry}, {Bamford}, {Hopkins}, {Loveday}, {Peacock},
  {Andrae}, {Bland-Hawthorn}, {Brough}, {Brown}, {Cameron}, {Ching}, {Colless},
  {Conselice}, {Croom}, {Cross}, {de Propris}, {Dye}, {Drinkwater}, {Ellis},
  {Graham}, {Grootes}, {Gunawardhana}, {Jones}, {van Kampen}, {Maraston},
  {Nichol}, {Parkinson}, {Phillipps}, {Pimbblet}, {Popescu}, {Prescott},
  {Roseboom}, {Sadler}, {Sansom}, {Sharp}, {Smith}, {Taylor}, {Thomas},
  {Tuffs}, {Wijesinghe}, {Dunne}, {Frenk}, {Jarvis}, {Madore}, {Meyer},
  {Seibert}, {Staveley-Smith}, {Sutherland}, \& {Warren}}]{Driver2011}
{Driver}, S.~P., {Hill}, D.~T., {Kelvin}, L.~S., {et~al.} 2011, \mnras, 413,
  971

\bibitem[{{Eddington}(1913)}]{Eddington1913}
{Eddington}, A.~S. 1913, \mnras, 73, 359

\bibitem[{{Emerson} {et~al.}(2004){Emerson}, {Sutherland}, {McPherson},
  {Craig}, {Dalton}, \& {Ward}}]{Emerson2004}
{Emerson}, J.~P., {Sutherland}, W.~J., {McPherson}, A.~M., {et~al.} 2004, The
  Messenger, 117, 27

\bibitem[{{Euclid Collaboration} {et~al.}(2022){Euclid Collaboration},
  {Moneti}, {McCracken}, {Shuntov}, {Kauffmann}, {Capak}, {Davidzon}, {Ilbert},
  {Scarlata}, {Toft}, {Weaver}, {Chary}, {Cuby}, {Faisst}, {Masters},
  {McPartland}, {Mobasher}, {Sanders}, {Scaramella}, {Stern}, {Szapudi},
  {Teplitz}, {Zalesky}, {Amara}, {Auricchio}, {Bodendorf}, {Bonino},
  {Branchini}, {Brau-Nogue}, {Brescia}, {Brinchmann}, {Capobianco}, {Carbone},
  {Carretero}, {Castander}, {Castellano}, {Cavuoti}, {Cimatti}, {Cledassou},
  {Congedo}, {Conselice}, {Conversi}, {Copin}, {Corcione}, {Costille},
  {Cropper}, {Da Silva}, {Degaudenzi}, {Douspis}, {Dubath}, {Duncan}, {Dupac},
  {Dusini}, {Farrens}, {Ferriol}, {Fosalba}, {Frailis}, {Franceschi}, {Fumana},
  {Garilli}, {Gillis}, {Giocoli}, {Granett}, {Grazian}, {Grupp}, {Haugan},
  {Hoekstra}, {Holmes}, {Hormuth}, {Hudelot}, {Jahnke}, {Kermiche},
  {Kiessling}, {Kilbinger}, {Kitching}, {Kohley}, {K{\"u}mmel}, {Kunz},
  {Kurki-Suonio}, {Ligori}, {Lilje}, {Lloro}, {Maiorano}, {Mansutti},
  {Marggraf}, {Markovic}, {Marulli}, {Massey}, {Maurogordato}, {Meneghetti},
  {Merlin}, {Meylan}, {Moresco}, {Moscardini}, {Munari}, {Niemi}, {Padilla},
  {Paltani}, {Pasian}, {Pedersen}, {Pires}, {Poncet}, {Popa}, {Pozzetti},
  {Raison}, {Rebolo}, {Rhodes}, {Rix}, {Roncarelli}, {Rossetti}, {Saglia},
  {Schneider}, {Secroun}, {Seidel}, {Serrano}, {Sirignano}, {Sirri}, {Stanco},
  {Tallada-Cresp{\'\i}}, {Taylor}, {Tereno}, {Toledo-Moreo}, {Torradeflot},
  {Wang}, {Welikala}, {Weller}, {Zamorani}, {Zoubian}, {Andreon}, {Bardelli},
  {Camera}, {Graci{\'a}-Carpio}, {Medinaceli}, {Mei}, {Polenta}, {Romelli},
  {Sureau}, {Tenti}, {Vassallo}, {Zacchei}, {Zucca}, {Baccigalupi},
  {Balaguera-Antol{\'\i}nez}, {Bernardeau}, {Biviano}, {Bolzonella}, {Bozzo},
  {Burigana}, {Cabanac}, {Cappi}, {Carvalho}, {Casas}, {Castignani},
  {Colodro-Conde}, {Coupon}, {Courtois}, {Di Ferdinando}, {Farina}, {Finelli},
  {Flose-Reimberg}, {Fotopoulou}, {Galeotta}, {Ganga}, {Garcia-Bellido},
  {Gaztanaga}, {Gozaliasl}, {Hook}, {Joachimi}, {Kansal}, {Keihanen},
  {Kirkpatrick}, {Lindholm}, {Mainetti}, {Maino}, {Maoli}, {Martinelli},
  {Martinet}, {Maturi}, {Metcalf}, {Morgante}, {Morisset}, {Nucita},
  {Patrizii}, {Potter}, {Renzi}, {Riccio}, {S{\'a}nchez}, {Sapone}, {Schirmer},
  {Schultheis}, {Scottez}, {Sefusatti}, {Teyssier}, {Tubio}, {Tutusaus},
  {Valiviita}, {Viel}, \& {Hildebrandt}}]{Moneti2022}
{Euclid Collaboration}, {Moneti}, A., {McCracken}, H.~J., {et~al.} 2022, \aap,
  658, A126

\bibitem[{{Felten}(1977)}]{Felten1977}
{Felten}, J.~E. 1977, \aj, 82, 861

\bibitem[{{Fernandez-Conde} {et~al.}(2008){Fernandez-Conde}, {Lagache},
  {Puget}, \& {Dole}}]{FC2008}
{Fernandez-Conde}, N., {Lagache}, G., {Puget}, J.~L., \& {Dole}, H. 2008, \aap,
  481, 885

\bibitem[{{Gaia Collaboration} {et~al.}(2016){Gaia Collaboration}, {Prusti,
  T.}, {de Bruijne, J. H. J.}, {Brown, A. G. A.}, {Vallenari, A.}, {Babusiaux,
  C.}, {Bailer-Jones, C. A. L.}, {Bastian, U.}, {Biermann, M.}, {Evans, D. W.},
  {Eyer, L.}, {Jansen, F.}, {Jordi, C.}, {Klioner, S. A.}, {Lammers, U.},
  {Lindegren, L.}, {Luri, X.}, {Mignard, F.}, {Milligan, D. J.}, {Panem, C.},
  {Poinsignon, V.}, {Pourbaix, D.}, {Randich, S.}, {Sarri, G.}, {Sartoretti,
  P.}, {Siddiqui, H. I.}, {Soubiran, C.}, {Valette, V.}, {van Leeuwen, F.},
  {Walton, N. A.}, {Aerts, C.}, {Arenou, F.}, {Cropper, M.}, {Drimmel, R.},
  {Høg, E.}, {Katz, D.}, {Lattanzi, M. G.}, {O’Mullane, W.}, {Grebel, E.
  K.}, {Holland, A. D.}, {Huc, C.}, {Passot, X.}, {Bramante, L.}, {Cacciari,
  C.}, {Castañeda, J.}, {Chaoul, L.}, {Cheek, N.}, {De Angeli, F.},
  {Fabricius, C.}, {Guerra, R.}, {Hernández, J.}, {Jean-Antoine-Piccolo, A.},
  {Masana, E.}, {Messineo, R.}, {Mowlavi, N.}, {Nienartowicz, K.},
  {Ordóñez-Blanco, D.}, {Panuzzo, P.}, {Portell, J.}, {Richards, P. J.},
  {Riello, M.}, {Seabroke, G. M.}, {Tanga, P.}, {Thévenin, F.}, {Torra, J.},
  {Els, S. G.}, {Gracia-Abril, G.}, {Comoretto, G.}, {Garcia-Reinaldos, M.},
  {Lock, T.}, {Mercier, E.}, {Altmann, M.}, {Andrae, R.}, {Astraatmadja, T.
  L.}, {Bellas-Velidis, I.}, {Benson, K.}, {Berthier, J.}, {Blomme, R.},
  {Busso, G.}, {Carry, B.}, {Cellino, A.}, {Clementini, G.}, {Cowell, S.},
  {Creevey, O.}, {Cuypers, J.}, {Davidson, M.}, {De Ridder, J.}, {de Torres,
  A.}, {Delchambre, L.}, {Dell’Oro, A.}, {Ducourant, C.}, {Frémat, Y.},
  {García-Torres, M.}, {Gosset, E.}, {Halbwachs, J.-L.}, {Hambly, N. C.},
  {Harrison, D. L.}, {Hauser, M.}, {Hestroffer, D.}, {Hodgkin, S. T.}, {Huckle,
  H. E.}, {Hutton, A.}, {Jasniewicz, G.}, {Jordan, S.}, {Kontizas, M.}, {Korn,
  A. J.}, {Lanzafame, A. C.}, {Manteiga, M.}, {Moitinho, A.}, {Muinonen, K.},
  {Osinde, J.}, {Pancino, E.}, {Pauwels, T.}, {Petit, J.-M.}, {Recio-Blanco,
  A.}, {Robin, A. C.}, {Sarro, L. M.}, {Siopis, C.}, {Smith, M.}, {Smith, K.
  W.}, {Sozzetti, A.}, {Thuillot, W.}, {van Reeven, W.}, {Viala, Y.}, {Abbas,
  U.}, {Abreu Aramburu, A.}, {Accart, S.}, {Aguado, J. J.}, {Allan, P. M.},
  {Allasia, W.}, {Altavilla, G.}, {Álvarez, M. A.}, {Alves, J.}, {Anderson, R.
  I.}, {Andrei, A. H.}, {Anglada Varela, E.}, {Antiche, E.}, {Antoja, T.},
  {Antón, S.}, {Arcay, B.}, {Atzei, A.}, {Ayache, L.}, {Bach, N.}, {Baker, S.
  G.}, {Balaguer-Núñez, L.}, {Barache, C.}, {Barata, C.}, {Barbier, A.},
  {Barblan, F.}, {Baroni, M.}, {Barrado y Navascués, D.}, {Barros, M.},
  {Barstow, M. A.}, {Becciani, U.}, {Bellazzini, M.}, {Bellei, G.}, {Bello
  García, A.}, {Belokurov, V.}, {Bendjoya, P.}, {Berihuete, A.}, {Bianchi,
  L.}, {Bienaymé, O.}, {Billebaud, F.}, {Blagorodnova, N.}, {Blanco-Cuaresma,
  S.}, {Boch, T.}, {Bombrun, A.}, {Borrachero, R.}, {Bouquillon, S.}, {Bourda,
  G.}, {Bouy, H.}, {Bragaglia, A.}, {Breddels, M. A.}, {Brouillet, N.},
  {Brüsemeister, T.}, {Bucciarelli, B.}, {Budnik, F.}, {Burgess, P.}, {Burgon,
  R.}, {Burlacu, A.}, {Busonero, D.}, {Buzzi, R.}, {Caffau, E.}, {Cambras, J.},
  {Campbell, H.}, {Cancelliere, R.}, {Cantat-Gaudin, T.}, {Carlucci, T.},
  {Carrasco, J. M.}, {Castellani, M.}, {Charlot, P.}, {Charnas, J.}, {Charvet,
  P.}, {Chassat, F.}, {Chiavassa, A.}, {Clotet, M.}, {Cocozza, G.}, {Collins,
  R. S.}, {Collins, P.}, {Costigan, G.}, {Crifo, F.}, {Cross, N. J. G.},
  {Crosta, M.}, {Crowley, C.}, {Dafonte, C.}, {Damerdji, Y.}, {Dapergolas, A.},
  {David, P.}, {David, M.}, {De Cat, P.}, {de Felice, F.}, {de Laverny, P.},
  {De Luise, F.}, {De March, R.}, {de Martino, D.}, {de Souza, R.},
  {Debosscher, J.}, {del Pozo, E.}, {Delbo, M.}, {Delgado, A.}, {Delgado, H.
  E.}, {di Marco, F.}, {Di Matteo, P.}, {Diakite, S.}, {Distefano, E.},
  {Dolding, C.}, {Dos Anjos, S.}, {Drazinos, P.}, {Durán, J.}, {Dzigan, Y.},
  {Ecale, E.}, {Edvardsson, B.}, {Enke, H.}, {Erdmann, M.}, {Escolar, D.},
  {Espina, M.}, {Evans, N. W.}, {Eynard Bontemps, G.}, {Fabre, C.}, {Fabrizio,
  M.}, {Faigler, S.}, {Falcão, A. J.}, {Farràs Casas, M.}, {Faye, F.},
  {Federici, L.}, {Fedorets, G.}, {Fernández-Hernández, J.}, {Fernique, P.},
  {Fienga, A.}, {Figueras, F.}, {Filippi, F.}, {Findeisen, K.}, {Fonti, A.},
  {Fouesneau, M.}, {Fraile, E.}, {Fraser, M.}, {Fuchs, J.}, {Furnell, R.},
  {Gai, M.}, {Galleti, S.}, {Galluccio, L.}, {Garabato, D.}, {García-Sedano,
  F.}, {Garé, P.}, {Garofalo, A.}, {Garralda, N.}, {Gavras, P.}, {Gerssen,
  J.}, {Geyer, R.}, {Gilmore, G.}, {Girona, S.}, {Giuffrida, G.}, {Gomes, M.},
  {González-Marcos, A.}, {González-Núñez, J.}, {González-Vidal, J. J.},
  {Granvik, M.}, {Guerrier, A.}, {Guillout, P.}, {Guiraud, J.}, {Gúrpide, A.},
  {Gutiérrez-Sánchez, R.}, {Guy, L. P.}, {Haigron, R.}, {Hatzidimitriou, D.},
  {Haywood, M.}, {Heiter, U.}, {Helmi, A.}, {Hobbs, D.}, {Hofmann, W.}, {Holl,
  B.}, {Holland, G.}, {Hunt, J. A. S.}, {Hypki, A.}, {Icardi, V.}, {Irwin, M.},
  {Jevardat de Fombelle, G.}, {Jofré, P.}, {Jonker, P. G.}, {Jorissen, A.},
  {Julbe, F.}, {Karampelas, A.}, {Kochoska, A.}, {Kohley, R.}, {Kolenberg, K.},
  {Kontizas, E.}, {Koposov, S. E.}, {Kordopatis, G.}, {Koubsky, P.},
  {Kowalczyk, A.}, {Krone-Martins, A.}, {Kudryashova, M.}, {Kull, I.},
  {Bachchan, R. K.}, {Lacoste-Seris, F.}, {Lanza, A. F.}, {Lavigne, J.-B.}, {Le
  Poncin-Lafitte, C.}, {Lebreton, Y.}, {Lebzelter, T.}, {Leccia, S.}, {Leclerc,
  N.}, {Lecoeur-Taibi, I.}, {Lemaitre, V.}, {Lenhardt, H.}, {Leroux, F.},
  {Liao, S.}, {Licata, E.}, {Lindstrøm, H. E. P.}, {Lister, T. A.}, {Livanou,
  E.}, {Lobel, A.}, {Löffler, W.}, {López, M.}, {Lopez-Lozano, A.}, {Lorenz,
  D.}, {Loureiro, T.}, {MacDonald, I.}, {Magalhães Fernandes, T.}, {Managau,
  S.}, {Mann, R. G.}, {Mantelet, G.}, {Marchal, O.}, {Marchant, J. M.},
  {Marconi, M.}, {Marie, J.}, {Marinoni, S.}, {Marrese, P. M.}, {Marschalkó,
  G.}, {Marshall, D. J.}, {Martín-Fleitas, J. M.}, {Martino, M.}, {Mary, N.},
  {Matijevič, G.}, {Mazeh, T.}, {McMillan, P. J.}, {Messina, S.}, {Mestre,
  A.}, {Michalik, D.}, {Millar, N. R.}, {Miranda, B. M. H.}, {Molina, D.},
  {Molinaro, R.}, {Molinaro, M.}, {Molnár, L.}, {Moniez, M.}, {Montegriffo,
  P.}, {Monteiro, D.}, {Mor, R.}, {Mora, A.}, {Morbidelli, R.}, {Morel, T.},
  {Morgenthaler, S.}, {Morley, T.}, {Morris, D.}, {Mulone, A. F.}, {Muraveva,
  T.}, {Musella, I.}, {Narbonne, J.}, {Nelemans, G.}, {Nicastro, L.}, {Noval,
  L.}, {Ordénovic, C.}, {Ordieres-Meré, J.}, {Osborne, P.}, {Pagani, C.},
  {Pagano, I.}, {Pailler, F.}, {Palacin, H.}, {Palaversa, L.}, {Parsons, P.},
  {Paulsen, T.}, {Pecoraro, M.}, {Pedrosa, R.}, {Pentikäinen, H.}, {Pereira,
  J.}, {Pichon, B.}, {Piersimoni, A. M.}, {Pineau, F.-X.}, {Plachy, E.}, {Plum,
  G.}, {Poujoulet, E.}, {Prša, A.}, {Pulone, L.}, {Ragaini, S.}, {Rago, S.},
  {Rambaux, N.}, {Ramos-Lerate, M.}, {Ranalli, P.}, {Rauw, G.}, {Read, A.},
  {Regibo, S.}, {Renk, F.}, {Reylé, C.}, {Ribeiro, R. A.}, {Rimoldini, L.},
  {Ripepi, V.}, {Riva, A.}, {Rixon, G.}, {Roelens, M.}, {Romero-Gómez, M.},
  {Rowell, N.}, {Royer, F.}, {Rudolph, A.}, {Ruiz-Dern, L.}, {Sadowski, G.},
  {Sagristà Sellés, T.}, {Sahlmann, J.}, {Salgado, J.}, {Salguero, E.},
  {Sarasso, M.}, {Savietto, H.}, {Schnorhk, A.}, {Schultheis, M.}, {Sciacca,
  E.}, {Segol, M.}, {Segovia, J. C.}, {Segransan, D.}, {Serpell, E.}, {Shih,
  I-C.}, {Smareglia, R.}, {Smart, R. L.}, {Smith, C.}, {Solano, E.}, {Solitro,
  F.}, {Sordo, R.}, {Soria Nieto, S.}, {Souchay, J.}, {Spagna, A.}, {Spoto,
  F.}, {Stampa, U.}, {Steele, I. A.}, {Steidelmüller, H.}, {Stephenson, C.
  A.}, {Stoev, H.}, {Suess, F. F.}, {Süveges, M.}, {Surdej, J.}, {Szabados,
  L.}, {Szegedi-Elek, E.}, {Tapiador, D.}, {Taris, F.}, {Tauran, G.}, {Taylor,
  M. B.}, {Teixeira, R.}, {Terrett, D.}, {Tingley, B.}, {Trager, S. C.},
  {Turon, C.}, {Ulla, A.}, {Utrilla, E.}, {Valentini, G.}, {van Elteren, A.},
  {Van Hemelryck, E.}, {van Leeuwen, M.}, {Varadi, M.}, {Vecchiato, A.},
  {Veljanoski, J.}, {Via, T.}, {Vicente, D.}, {Vogt, S.}, {Voss, H.}, {Votruba,
  V.}, {Voutsinas, S.}, {Walmsley, G.}, {Weiler, M.}, {Weingrill, K.}, {Werner,
  D.}, {Wevers, T.}, {Whitehead, G.}, {Wyrzykowski, Ł.}, {Yoldas, A.},
  {Žerjal, M.}, {Zucker, S.}, {Zurbach, C.}, {Zwitter, T.}, {Alecu, A.},
  {Allen, M.}, {Allende Prieto, C.}, {Amorim, A.}, {Anglada-Escudé, G.},
  {Arsenijevic, V.}, {Azaz, S.}, {Balm, P.}, {Beck, M.}, {Bernstein, H.-H.},
  {Bigot, L.}, {Bijaoui, A.}, {Blasco, C.}, {Bonfigli, M.}, {Bono, G.},
  {Boudreault, S.}, {Bressan, A.}, {Brown, S.}, {Brunet, P.-M.}, {Bunclark,
  P.}, {Buonanno, R.}, {Butkevich, A. G.}, {Carret, C.}, {Carrion, C.},
  {Chemin, L.}, {Chéreau, F.}, {Corcione, L.}, {Darmigny, E.}, {de Boer, K.
  S.}, {de Teodoro, P.}, {de Zeeuw, P. T.}, {Delle Luche, C.}, {Domingues, C.
  D.}, {Dubath, P.}, {Fodor, F.}, {Frézouls, B.}, {Fries, A.}, {Fustes, D.},
  {Fyfe, D.}, {Gallardo, E.}, {Gallegos, J.}, {Gardiol, D.}, {Gebran, M.},
  {Gomboc, A.}, {Gómez, A.}, {Grux, E.}, {Gueguen, A.}, {Heyrovsky, A.},
  {Hoar, J.}, {Iannicola, G.}, {Isasi Parache, Y.}, {Janotto, A.-M.}, {Joliet,
  E.}, {Jonckheere, A.}, {Keil, R.}, {Kim, D.-W.}, {Klagyivik, P.}, {Klar, J.},
  {Knude, J.}, {Kochukhov, O.}, {Kolka, I.}, {Kos, J.}, {Kutka, A.}, {Lainey,
  V.}, {LeBouquin, D.}, {Liu, C.}, {Loreggia, D.}, {Makarov, V. V.},
  {Marseille, M. G.}, {Martayan, C.}, {Martinez-Rubi, O.}, {Massart, B.},
  {Meynadier, F.}, {Mignot, S.}, {Munari, U.}, {Nguyen, A.-T.}, {Nordlander,
  T.}, {Ocvirk, P.}, {O’Flaherty, K. S.}, {Olias Sanz, A.}, {Ortiz, P.},
  {Osorio, J.}, {Oszkiewicz, D.}, {Ouzounis, A.}, {Palmer, M.}, {Park, P.},
  {Pasquato, E.}, {Peltzer, C.}, {Peralta, J.}, {Péturaud, F.}, {Pieniluoma,
  T.}, {Pigozzi, E.}, {Poels, J.}, {Prat, G.}, {Prod’homme, T.}, {Raison,
  F.}, {Rebordao, J. M.}, {Risquez, D.}, {Rocca-Volmerange, B.}, {Rosen, S.},
  {Ruiz-Fuertes, M. I.}, {Russo, F.}, {Sembay, S.}, {Serraller Vizcaino, I.},
  {Short, A.}, {Siebert, A.}, {Silva, H.}, {Sinachopoulos, D.}, {Slezak, E.},
  {Soffel, M.}, {Sosnowska, D.}, {Straižys, V.}, {ter Linden, M.}, {Terrell,
  D.}, {Theil, S.}, {Tiede, C.}, {Troisi, L.}, {Tsalmantza, P.}, {Tur, D.},
  {Vaccari, M.}, {Vachier, F.}, {Valles, P.}, {Van Hamme, W.}, {Veltz, L.},
  {Virtanen, J.}, {Wallut, J.-M.}, {Wichmann, R.}, {Wilkinson, M. I.},
  {Ziaeepour, H.}, \& {Zschocke, S.}}]{Gaia2016}
{Gaia Collaboration}, {Prusti, T.}, {de Bruijne, J. H. J.}, {et~al.} 2016,
  A\&A, 595, A1

\bibitem[{{Gaia Collaboration} {et~al.}(2018){Gaia Collaboration}, {Brown},
  {Vallenari}, {Prusti}, {de Bruijne}, {Babusiaux}, {Bailer-Jones}, {Biermann},
  {Evans}, {Eyer}, {Jansen}, {Jordi}, {Klioner}, {Lammers}, {Lindegren},
  {Luri}, {Mignard}, {Panem}, {Pourbaix}, {Randich}, {Sartoretti}, {Siddiqui},
  {Soubiran}, {van Leeuwen}, {Walton}, {Arenou}, {Bastian}, {Cropper},
  {Drimmel}, {Katz}, {Lattanzi}, {Bakker}, {Cacciari}, {Casta{\~n}eda},
  {Chaoul}, {Cheek}, {De Angeli}, {Fabricius}, {Guerra}, {Holl}, {Masana},
  {Messineo}, {Mowlavi}, {Nienartowicz}, {Panuzzo}, {Portell}, {Riello},
  {Seabroke}, {Tanga}, {Th{\'e}venin}, {Gracia-Abril}, {Comoretto},
  {Garcia-Reinaldos}, {Teyssier}, {Altmann}, {Andrae}, {Audard},
  {Bellas-Velidis}, {Benson}, {Berthier}, {Blomme}, {Burgess}, {Busso},
  {Carry}, {Cellino}, {Clementini}, {Clotet}, {Creevey}, {Davidson}, {De
  Ridder}, {Delchambre}, {Dell'Oro}, {Ducourant},
  {Fern{\'a}ndez-Hern{\'a}ndez}, {Fouesneau}, {Fr{\'e}mat}, {Galluccio},
  {Garc{\'\i}a-Torres}, {Gonz{\'a}lez-N{\'u}{\~n}ez}, {Gonz{\'a}lez-Vidal},
  {Gosset}, {Guy}, {Halbwachs}, {Hambly}, {Harrison}, {Hern{\'a}ndez},
  {Hestroffer}, {Hodgkin}, {Hutton}, {Jasniewicz}, {Jean-Antoine-Piccolo},
  {Jordan}, {Korn}, {Krone-Martins}, {Lanzafame}, {Lebzelter}, {L{\"o}ffler},
  {Manteiga}, {Marrese}, {Mart{\'\i}n-Fleitas}, {Moitinho}, {Mora}, {Muinonen},
  {Osinde}, {Pancino}, {Pauwels}, {Petit}, {Recio-Blanco}, {Richards},
  {Rimoldini}, {Robin}, {Sarro}, {Siopis}, {Smith}, {Sozzetti}, {S{\"u}veges},
  {Torra}, {van Reeven}, {Abbas}, {Abreu Aramburu}, {Accart}, {Aerts},
  {Altavilla}, {{\'A}lvarez}, {Alvarez}, {Alves}, {Anderson}, {Andrei},
  {Anglada Varela}, {Antiche}, {Antoja}, {Arcay}, {Astraatmadja}, {Bach},
  {Baker}, {Balaguer-N{\'u}{\~n}ez}, {Balm}, {Barache}, {Barata}, {Barbato},
  {Barblan}, {Barklem}, {Barrado}, {Barros}, {Barstow}, {Bartholom{\'e}
  Mu{\~n}oz}, {Bassilana}, {Becciani}, {Bellazzini}, {Berihuete}, {Bertone},
  {Bianchi}, {Bienaym{\'e}}, {Blanco-Cuaresma}, {Boch}, {Boeche}, {Bombrun},
  {Borrachero}, {Bossini}, {Bouquillon}, {Bourda}, {Bragaglia}, {Bramante},
  {Breddels}, {Bressan}, {Brouillet}, {Br{\"u}semeister}, {Brugaletta},
  {Bucciarelli}, {Burlacu}, {Busonero}, {Butkevich}, {Buzzi}, {Caffau},
  {Cancelliere}, {Cannizzaro}, {Cantat-Gaudin}, {Carballo}, {Carlucci},
  {Carrasco}, {Casamiquela}, {Castellani}, {Castro-Ginard}, {Charlot},
  {Chemin}, {Chiavassa}, {Cocozza}, {Costigan}, {Cowell}, {Crifo}, {Crosta},
  {Crowley}, {Cuypers}, {Dafonte}, {Damerdji}, {Dapergolas}, {David}, {David},
  {de Laverny}, {De Luise}, {De March}, {de Martino}, {de Souza}, {de Torres},
  {Debosscher}, {del Pozo}, {Delbo}, {Delgado}, {Delgado}, {Di Matteo},
  {Diakite}, {Diener}, {Distefano}, {Dolding}, {Drazinos}, {Dur{\'a}n},
  {Edvardsson}, {Enke}, {Eriksson}, {Esquej}, {Eynard Bontemps}, {Fabre},
  {Fabrizio}, {Faigler}, {Falc{\~a}o}, {Farr{\`a}s Casas}, {Federici},
  {Fedorets}, {Fernique}, {Figueras}, {Filippi}, {Findeisen}, {Fonti},
  {Fraile}, {Fraser}, {Fr{\'e}zouls}, {Gai}, {Galleti}, {Garabato},
  {Garc{\'\i}a-Sedano}, {Garofalo}, {Garralda}, {Gavel}, {Gavras}, {Gerssen},
  {Geyer}, {Giacobbe}, {Gilmore}, {Girona}, {Giuffrida}, {Glass}, {Gomes},
  {Granvik}, {Gueguen}, {Guerrier}, {Guiraud}, {Guti{\'e}rrez-S{\'a}nchez},
  {Haigron}, {Hatzidimitriou}, {Hauser}, {Haywood}, {Heiter}, {Helmi}, {Heu},
  {Hilger}, {Hobbs}, {Hofmann}, {Holland}, {Huckle}, {Hypki}, {Icardi},
  {Jan{\ss}en}, {Jevardat de Fombelle}, {Jonker}, {Juh{\'a}sz}, {Julbe},
  {Karampelas}, {Kewley}, {Klar}, {Kochoska}, {Kohley}, {Kolenberg},
  {Kontizas}, {Kontizas}, {Koposov}, {Kordopatis}, {Kostrzewa-Rutkowska},
  {Koubsky}, {Lambert}, {Lanza}, {Lasne}, {Lavigne}, {Le Fustec}, {Le
  Poncin-Lafitte}, {Lebreton}, {Leccia}, {Leclerc}, {Lecoeur-Taibi},
  {Lenhardt}, {Leroux}, {Liao}, {Licata}, {Lindstr{\o}m}, {Lister}, {Livanou},
  {Lobel}, {L{\'o}pez}, {Managau}, {Mann}, {Mantelet}, {Marchal}, {Marchant},
  {Marconi}, {Marinoni}, {Marschalk{\'o}}, {Marshall}, {Martino}, {Marton},
  {Mary}, {Massari}, {Matijevi{\v{c}}}, {Mazeh}, {McMillan}, {Messina},
  {Michalik}, {Millar}, {Molina}, {Molinaro}, {Moln{\'a}r}, {Montegriffo},
  {Mor}, {Morbidelli}, {Morel}, {Morris}, {Mulone}, {Muraveva}, {Musella},
  {Nelemans}, {Nicastro}, {Noval}, {O'Mullane}, {Ord{\'e}novic},
  {Ord{\'o}{\~n}ez-Blanco}, {Osborne}, {Pagani}, {Pagano}, {Pailler},
  {Palacin}, {Palaversa}, {Panahi}, {Pawlak}, {Piersimoni}, {Pineau}, {Plachy},
  {Plum}, {Poggio}, {Poujoulet}, {Pr{\v{s}}a}, {Pulone}, {Racero}, {Ragaini},
  {Rambaux}, {Ramos-Lerate}, {Regibo}, {Reyl{\'e}}, {Riclet}, {Ripepi}, {Riva},
  {Rivard}, {Rixon}, {Roegiers}, {Roelens}, {Romero-G{\'o}mez}, {Rowell},
  {Royer}, {Ruiz-Dern}, {Sadowski}, {Sagrist{\`a} Sell{\'e}s}, {Sahlmann},
  {Salgado}, {Salguero}, {Sanna}, {Santana-Ros}, {Sarasso}, {Savietto},
  {Schultheis}, {Sciacca}, {Segol}, {Segovia}, {S{\'e}gransan}, {Shih},
  {Siltala}, {Silva}, {Smart}, {Smith}, {Solano}, {Solitro}, {Sordo}, {Soria
  Nieto}, {Souchay}, {Spagna}, {Spoto}, {Stampa}, {Steele},
  {Steidelm{\"u}ller}, {Stephenson}, {Stoev}, {Suess}, {Surdej}, {Szabados},
  {Szegedi-Elek}, {Tapiador}, {Taris}, {Tauran}, {Taylor}, {Teixeira},
  {Terrett}, {Teyssandier}, {Thuillot}, {Titarenko}, {Torra Clotet}, {Turon},
  {Ulla}, {Utrilla}, {Uzzi}, {Vaillant}, {Valentini}, {Valette}, {van Elteren},
  {Van Hemelryck}, {van Leeuwen}, {Vaschetto}, {Vecchiato}, {Veljanoski},
  {Viala}, {Vicente}, {Vogt}, {von Essen}, {Voss}, {Votruba}, {Voutsinas},
  {Walmsley}, {Weiler}, {Wertz}, {Wevers}, {Wyrzykowski}, {Yoldas},
  {{\v{Z}}erjal}, {Ziaeepour}, {Zorec}, {Zschocke}, {Zucker}, {Zurbach}, \&
  {Zwitter}}]{Gaia2018}
{Gaia Collaboration}, {Brown}, A.~G.~A., {Vallenari}, A., {et~al.} 2018, \aap,
  616, A1

\bibitem[{{Grogin} {et~al.}(2011){Grogin}, {Kocevski}, {Faber}, {Ferguson},
  {Koekemoer}, {Riess}, {Acquaviva}, {Alexander}, {Almaini}, {Ashby}, {Barden},
  {Bell}, {Bournaud}, {Brown}, {Caputi}, {Casertano}, {Cassata}, {Castellano},
  {Challis}, {Chary}, {Cheung}, {Cirasuolo}, {Conselice}, {Roshan Cooray},
  {Croton}, {Daddi}, {Dahlen}, {Dav{\'e}}, {de Mello}, {Dekel}, {Dickinson},
  {Dolch}, {Donley}, {Dunlop}, {Dutton}, {Elbaz}, {Fazio}, {Filippenko},
  {Finkelstein}, {Fontana}, {Gardner}, {Garnavich}, {Gawiser}, {Giavalisco},
  {Grazian}, {Guo}, {Hathi}, {H{\"a}ussler}, {Hopkins}, {Huang}, {Huang},
  {Jha}, {Kartaltepe}, {Kirshner}, {Koo}, {Lai}, {Lee}, {Li}, {Lotz}, {Lucas},
  {Madau}, {McCarthy}, {McGrath}, {McIntosh}, {McLure}, {Mobasher},
  {Moustakas}, {Mozena}, {Nandra}, {Newman}, {Niemi}, {Noeske}, {Papovich},
  {Pentericci}, {Pope}, {Primack}, {Rajan}, {Ravindranath}, {Reddy}, {Renzini},
  {Rix}, {Robaina}, {Rodney}, {Rosario}, {Rosati}, {Salimbeni}, {Scarlata},
  {Siana}, {Simard}, {Smidt}, {Somerville}, {Spinrad}, {Straughn}, {Strolger},
  {Telford}, {Teplitz}, {Trump}, {van der Wel}, {Villforth}, {Wechsler},
  {Weiner}, {Wiklind}, {Wild}, {Wilson}, {Wuyts}, {Yan}, \& {Yun}}]{Grogin2011}
{Grogin}, N.~A., {Kocevski}, D.~D., {Faber}, S.~M., {et~al.} 2011, \apjs, 197,
  35

\bibitem[{{Jarvis} {et~al.}(2013){Jarvis}, {Bonfield}, {Bruce}, {Geach},
  {McAlpine}, {McLure}, {Gonz{\'a}lez-Solares}, {Irwin}, {Lewis}, {Yoldas},
  {Andreon}, {Cross}, {Emerson}, {Dalton}, {Dunlop}, {Hodgkin}, {Le},
  {Karouzos}, {Meisenheimer}, {Oliver}, {Rawlings}, {Simpson}, {Smail},
  {Smith}, {Sullivan}, {Sutherland}, {White}, \& {Zwart}}]{Jarvis2013}
{Jarvis}, M.~J., {Bonfield}, D.~G., {Bruce}, V.~A., {et~al.} 2013, \mnras, 428,
  1281

\bibitem[{{Klypin} {et~al.}(2016){Klypin}, {Yepes}, {Gottl{\"o}ber}, {Prada},
  \& {He{\ss}}}]{Klypin2016}
{Klypin}, A., {Yepes}, G., {Gottl{\"o}ber}, S., {Prada}, F., \& {He{\ss}}, S.
  2016, \mnras, 457, 4340

\bibitem[{{Laigle} {et~al.}(2016){Laigle}, {McCracken}, {Ilbert}, {Hsieh},
  {Davidzon}, {Capak}, {Hasinger}, {Silverman}, {Pichon}, {Coupon}, {Aussel},
  {Le Borgne}, {Caputi}, {Cassata}, {Chang}, {Civano}, {Dunlop}, {Fynbo},
  {Kartaltepe}, {Koekemoer}, {Le F{\`e}vre}, {Le Floc'h}, {Leauthaud}, {Lilly},
  {Lin}, {Marchesi}, {Milvang-Jensen}, {Salvato}, {Sanders}, {Scoville},
  {Smolcic}, {Stockmann}, {Taniguchi}, {Tasca}, {Toft}, {Vaccari}, \&
  {Zabl}}]{Laigle2016}
{Laigle}, C., {McCracken}, H.~J., {Ilbert}, O., {et~al.} 2016, \apjs, 224, 24

\bibitem[{{Lane} {et~al.}(2007){Lane}, {Almaini}, {Foucaud}, {Simpson},
  {Smail}, {McLure}, {Conselice}, {Cirasuolo}, {Page}, {Dunlop}, {Hirst},
  {Watson}, \& {Sekiguchi}}]{Lane2007}
{Lane}, K.~P., {Almaini}, O., {Foucaud}, S., {et~al.} 2007, \mnras, 379, L25

\bibitem[{{McCracken} {et~al.}(2012){McCracken}, {Milvang-Jensen}, {Dunlop},
  {Franx}, {Fynbo}, {Le F{\`e}vre}, {Holt}, {Caputi}, {Goranova}, {Buitrago},
  {Emerson}, {Freudling}, {Hudelot}, {L{\'o}pez-Sanjuan}, {Magnard}, {Mellier},
  {M{\o}ller}, {Nilsson}, {Sutherland}, {Tasca}, \& {Zabl}}]{McCracken2012}
{McCracken}, H.~J., {Milvang-Jensen}, B., {Dunlop}, J., {et~al.} 2012, \aap,
  544, A156

\bibitem[{{Moneti} {et~al.}(2023){Moneti}, {McCracken}, {Hudelot}, {Rouberol},
  {Herent}, {Mellier}, {Dunlop}, {Le Fevre}, {Franx}, {Fynbo}, {Bowler},
  {Caputi}, {Kauffmann}, {Milvang-Jensen}, {Gonzalez-Fernandez},
  {Gonzalez-Solares}, {Irwin}, {Lewis}, {Blake}, {Cross}, {Read}, \&
  {Sutorius}}]{Moneti2023}
{Moneti}, A., {McCracken}, H.~J., {Hudelot}, W., {et~al.} 2023, VizieR Online
  Data Catalog, II/373

\bibitem[{{Navarro} {et~al.}(1997){Navarro}, {Frenk}, \& {White}}]{NFW}
{Navarro}, J.~F., {Frenk}, C.~S., \& {White}, S. D.~M. 1997, \apj, 490, 493

\bibitem[{{Patanchon} {et~al.}(2009){Patanchon}, {Ade}, {Bock}, {Chapin},
  {Devlin}, {Dicker}, {Griffin}, {Gundersen}, {Halpern}, {Hargrave}, {Hughes},
  {Klein}, {Marsden}, {Mauskopf}, {Moncelsi}, {Netterfield}, {Olmi}, {Pascale},
  {Rex}, {Scott}, {Semisch}, {Thomas}, {Truch}, {Tucker}, {Tucker}, {Viero}, \&
  {Wiebe}}]{Patanchon2009}
{Patanchon}, G., {Ade}, P. A.~R., {Bock}, J.~J., {et~al.} 2009, \apj, 707, 1750

\bibitem[{{Planck Collaboration} {et~al.}(2020){Planck Collaboration},
  {Aghanim}, {Akrami}, {Ashdown}, {Aumont}, {Baccigalupi}, {Ballardini},
  {Banday}, {Barreiro}, {Bartolo}, {Basak}, {Battye}, {Benabed}, {Bernard},
  {Bersanelli}, {Bielewicz}, {Bock}, {Bond}, {Borrill}, {Bouchet}, {Boulanger},
  {Bucher}, {Burigana}, {Butler}, {Calabrese}, {Cardoso}, {Carron},
  {Challinor}, {Chiang}, {Chluba}, {Colombo}, {Combet}, {Contreras}, {Crill},
  {Cuttaia}, {de Bernardis}, {de Zotti}, {Delabrouille}, {Delouis}, {Di
  Valentino}, {Diego}, {Dor{\'e}}, {Douspis}, {Ducout}, {Dupac}, {Dusini},
  {Efstathiou}, {Elsner}, {En{\ss}lin}, {Eriksen}, {Fantaye}, {Farhang},
  {Fergusson}, {Fernandez-Cobos}, {Finelli}, {Forastieri}, {Frailis},
  {Fraisse}, {Franceschi}, {Frolov}, {Galeotta}, {Galli}, {Ganga},
  {G{\'e}nova-Santos}, {Gerbino}, {Ghosh}, {Gonz{\'a}lez-Nuevo}, {G{\'o}rski},
  {Gratton}, {Gruppuso}, {Gudmundsson}, {Hamann}, {Handley}, {Hansen},
  {Herranz}, {Hildebrandt}, {Hivon}, {Huang}, {Jaffe}, {Jones}, {Karakci},
  {Keih{\"a}nen}, {Keskitalo}, {Kiiveri}, {Kim}, {Kisner}, {Knox},
  {Krachmalnicoff}, {Kunz}, {Kurki-Suonio}, {Lagache}, {Lamarre}, {Lasenby},
  {Lattanzi}, {Lawrence}, {Le Jeune}, {Lemos}, {Lesgourgues}, {Levrier},
  {Lewis}, {Liguori}, {Lilje}, {Lilley}, {Lindholm}, {L{\'o}pez-Caniego},
  {Lubin}, {Ma}, {Mac{\'\i}as-P{\'e}rez}, {Maggio}, {Maino}, {Mandolesi},
  {Mangilli}, {Marcos-Caballero}, {Maris}, {Martin}, {Martinelli},
  {Mart{\'\i}nez-Gonz{\'a}lez}, {Matarrese}, {Mauri}, {McEwen}, {Meinhold},
  {Melchiorri}, {Mennella}, {Migliaccio}, {Millea}, {Mitra},
  {Miville-Desch{\^e}nes}, {Molinari}, {Montier}, {Morgante}, {Moss}, {Natoli},
  {N{\o}rgaard-Nielsen}, {Pagano}, {Paoletti}, {Partridge}, {Patanchon},
  {Peiris}, {Perrotta}, {Pettorino}, {Piacentini}, {Polastri}, {Polenta},
  {Puget}, {Rachen}, {Reinecke}, {Remazeilles}, {Renzi}, {Rocha}, {Rosset},
  {Roudier}, {Rubi{\~n}o-Mart{\'\i}n}, {Ruiz-Granados}, {Salvati}, {Sandri},
  {Savelainen}, {Scott}, {Shellard}, {Sirignano}, {Sirri}, {Spencer},
  {Sunyaev}, {Suur-Uski}, {Tauber}, {Tavagnacco}, {Tenti}, {Toffolatti},
  {Tomasi}, {Trombetti}, {Valenziano}, {Valiviita}, {Van Tent}, {Vibert},
  {Vielva}, {Villa}, {Vittorio}, {Wandelt}, {Wehus}, {White}, {White},
  {Zacchei}, \& {Zonca}}]{Planck2018}
{Planck Collaboration}, {Aghanim}, N., {Akrami}, Y., {et~al.} 2020, \aap, 641,
  A6

\bibitem[{{Roseboom} {et~al.}(2010){Roseboom}, {Oliver}, {Kunz}, {Altieri},
  {Amblard}, {Arumugam}, {Auld}, {Aussel}, {Babbedge}, {B{\'e}thermin},
  {Blain}, {Bock}, {Boselli}, {Brisbin}, {Buat}, {Burgarella},
  {Castro-Rodr{\'\i}guez}, {Cava}, {Chanial}, {Chapin}, {Clements}, {Conley},
  {Conversi}, {Cooray}, {Dowell}, {Dwek}, {Dye}, {Eales}, {Elbaz}, {Farrah},
  {Fox}, {Franceschini}, {Gear}, {Glenn}, {Solares}, {Griffin}, {Halpern},
  {Harwit}, {Hatziminaoglou}, {Huang}, {Ibar}, {Isaak}, {Ivison}, {Lagache},
  {Levenson}, {Lu}, {Madden}, {Maffei}, {Mainetti}, {Marchetti}, {Marsden},
  {Mortier}, {Nguyen}, {O'Halloran}, {Omont}, {Page}, {Panuzzo},
  {Papageorgiou}, {Patel}, {Pearson}, {P{\'e}rez-Fournon}, {Pohlen},
  {Rawlings}, {Raymond}, {Rigopoulou}, {Rizzo}, {Rowan-Robinson}, {Portal},
  {Schulz}, {Scott}, {Seymour}, {Shupe}, {Smith}, {Stevens}, {Symeonidis},
  {Trichas}, {Tugwell}, {Vaccari}, {Valtchanov}, {Vieira}, {Vigroux}, {Wang},
  {Ward}, {Wright}, {Xu}, \& {Zemcov}}]{Roseboom2010}
{Roseboom}, I.~G., {Oliver}, S.~J., {Kunz}, M., {et~al.} 2010, \mnras, 409, 48

\bibitem[{{Sawicki}(2002)}]{Sawicki_2002}
{Sawicki}, M. 2002, \aj, 124, 3050

\bibitem[{{Scoville} {et~al.}(2007){Scoville}, {Aussel}, {Brusa}, {Capak},
  {Carollo}, {Elvis}, {Giavalisco}, {Guzzo}, {Hasinger}, {Impey}, {Kneib},
  {LeFevre}, {Lilly}, {Mobasher}, {Renzini}, {Rich}, {Sanders}, {Schinnerer},
  {Schminovich}, {Shopbell}, {Taniguchi}, \& {Tyson}}]{Scoville_2007}
{Scoville}, N., {Aussel}, H., {Brusa}, M., {et~al.} 2007, \apjs, 172, 1

\bibitem[{{Somerville} {et~al.}(2008){Somerville}, {Hopkins}, {Cox},
  {Robertson}, \& {Hernquist}}]{Somerville2008}
{Somerville}, R.~S., {Hopkins}, P.~F., {Cox}, T.~J., {Robertson}, B.~E., \&
  {Hernquist}, L. 2008, \mnras, 391, 481

\bibitem[{{Somerville} {et~al.}(2015){Somerville}, {Popping}, \&
  {Trager}}]{Somerville2015}
{Somerville}, R.~S., {Popping}, G., \& {Trager}, S.~C. 2015, \mnras, 453, 4337

\bibitem[{{Somerville} {et~al.}(2021){Somerville}, {Olsen}, {Yung}, {Pacifici},
  {Ferguson}, {Behroozi}, {Osborne}, {Wechsler}, {Pandya}, {Faber}, {Primack},
  \& {Dekel}}]{Somerville2021}
{Somerville}, R.~S., {Olsen}, C., {Yung}, L.~Y.~A., {et~al.} 2021, \mnras, 502,
  4858

\bibitem[{{Spergel} {et~al.}(2015){Spergel}, {Gehrels}, {Baltay}, {Bennett},
  {Breckinridge}, {Donahue}, {Dressler}, {Gaudi}, {Greene}, {Guyon}, {Hirata},
  {Kalirai}, {Kasdin}, {Macintosh}, {Moos}, {Perlmutter}, {Postman},
  {Rauscher}, {Rhodes}, {Wang}, {Weinberg}, {Benford}, {Hudson}, {Jeong},
  {Mellier}, {Traub}, {Yamada}, {Capak}, {Colbert}, {Masters}, {Penny},
  {Savransky}, {Stern}, {Zimmerman}, {Barry}, {Bartusek}, {Carpenter}, {Cheng},
  {Content}, {Dekens}, {Demers}, {Grady}, {Jackson}, {Kuan}, {Kruk}, {Melton},
  {Nemati}, {Parvin}, {Poberezhskiy}, {Peddie}, {Ruffa}, {Wallace}, {Whipple},
  {Wollack}, \& {Zhao}}]{spergel2015}
{Spergel}, D., {Gehrels}, N., {Baltay}, C., {et~al.} 2015, arXiv e-prints,
  arXiv:1503.03757

\bibitem[{{Weaver} {et~al.}(2022){Weaver}, {Kauffmann}, {Ilbert}, {McCracken},
  {Moneti}, {Toft}, {Brammer}, {Shuntov}, {Davidzon}, {Hsieh}, {Laigle},
  {Anastasiou}, {Jespersen}, {Vinther}, {Capak}, {Casey}, {McPartland},
  {Milvang-Jensen}, {Mobasher}, {Sanders}, {Zalesky}, {Arnouts}, {Aussel},
  {Dunlop}, {Faisst}, {Franx}, {Furtak}, {Fynbo}, {Gould}, {Greve}, {Gwyn},
  {Kartaltepe}, {Kashino}, {Koekemoer}, {Kokorev}, {Le F{\`e}vre}, {Lilly},
  {Masters}, {Magdis}, {Mehta}, {Peng}, {Riechers}, {Salvato}, {Sawicki},
  {Scarlata}, {Scoville}, {Shirley}, {Silverman}, {Sneppen}, {Smolc̆i{\'c}},
  {Steinhardt}, {Stern}, {Tanaka}, {Taniguchi}, {Teplitz}, {Vaccari}, {Wang},
  \& {Zamorani}}]{COSMOS2020}
{Weaver}, J.~R., {Kauffmann}, O.~B., {Ilbert}, O., {et~al.} 2022, \apjs, 258,
  11

\bibitem[{{Wright} {et~al.}(2010){Wright}, {Eisenhardt}, {Mainzer}, {Ressler},
  {Cutri}, {Jarrett}, {Kirkpatrick}, {Padgett}, {McMillan}, {Skrutskie},
  {Stanford}, {Cohen}, {Walker}, {Mather}, {Leisawitz}, {Gautier}, {McLean},
  {Benford}, {Lonsdale}, {Blain}, {Mendez}, {Irace}, {Duval}, {Liu}, {Royer},
  {Heinrichsen}, {Howard}, {Shannon}, {Kendall}, {Walsh}, {Larsen}, {Cardon},
  {Schick}, {Schwalm}, {Abid}, {Fabinsky}, {Naes}, \& {Tsai}}]{Wright2010}
{Wright}, E.~L., {Eisenhardt}, P. R.~M., {Mainzer}, A.~K., {et~al.} 2010, \aj,
  140, 1868

\bibitem[{{Yung} {et~al.}(2022){Yung}, {Somerville}, {Ferguson}, {Finkelstein},
  {Gardner}, {Dav{\'e}}, {Bagley}, {Popping}, \& {Behroozi}}]{Yung2022}
{Yung}, L.~Y.~A., {Somerville}, R.~S., {Ferguson}, H.~C., {et~al.} 2022,
  \mnras, 515, 5416

\bibitem[{{Yung} {et~al.}(2023){Yung}, {Somerville}, {Finkelstein}, {Behroozi},
  {Dav{\'e}}, {Ferguson}, {Gardner}, {Popping}, {Malhotra}, {Papovich},
  {Rhoads}, {Bagley}, {Hirschmann}, \& {Koekemoer}}]{Yung2022b}
{Yung}, L.~Y.~A., {Somerville}, R.~S., {Finkelstein}, S.~L., {et~al.} 2023,
  \mnras, 519, 1578

\end{thebibliography}






\end{document}